\documentclass{aa}
\usepackage{amssymb}
\usepackage{graphics}
\usepackage{multirow}

\newcommand{\NVab}{N{\sevenrm~V}\,$\lambda\lambda$1238,1242}

\def\CIV{C\,{\sc iv}}

\def\CIVab{C{\sc iv}\,$\lambda\lambda$1548,1551}

\def\MgII{Mg\,{\sc ii}}

\def\MgIIab{Mg{\sc ii}\,$\lambda\lambda$2796,2803}
\def\MgI{Mg\,{\sc i}\,$\lambda$2852}
\def\OIIwave{[O\,{\sc ii}]\,$\lambda$3728}

\def\FeII{Fe\,{\sc ii}}
\def\fea{Fe\,{\sc ii}\,$\lambda$2600}
\def\feb{Fe\,{\sc ii}\,$\lambda$2586}
\def\fec{Fe\,{\sc ii}\,$\lambda$2382}
\def\fed{Fe\,{\sc ii}\,$\lambda$2374}
\def\fee{Fe\,{\sc ii}\,$\lambda$2344}

\def\OII{[O\,{\sc ii}]}

\def\zabs{$z_{\rm abs}$}
\def\zem{$z_{\rm em}$}
\def\kms{$\rm km\,s^{-1}$}
\def\ergs{${\rm erg\,s^{-1}}$}

\def\aap{A\&Ap\/}

\def\apjs{ApJS}

\def\apjl{ApJL}

 \font\sevenrm=cmr7 scaled 1000

\begin{document}
\titlerunning{The inflowing \MgII\ NALs}
\title{The detections of inflowing gas from narrow absorption lines at the parsec scale}
\author{ Zhi-Fu Chen\inst{1,4,6}$^\ast$ \and Minfeng Gu\inst{2}$^\ast$  \and Zhicheng He\inst{3}$^\ast$ \and
Defu Bu\inst{2} \and Fulai Guo\inst{2}  \and Qiusheng Gu\inst{4} \and Yiping Qin\inst{5}}
\institute{
{Department of physics, Guangxi University for Nationalities, Nanning 530006, China; zhichenfu@126.com}
\and{Key Laboratory for Research in Galaxies and Cosmology, Shanghai Astronomical Observatory, Chinese Academy of Sciences, Shanghai 200030, China; gumf@shao.ac.cn}
\and {School of Astronomy and Space Science, University of Science and Technology of China, Hefei, China; zcho@ustc.edu.cn}
 \and {School of Astronomy and Space Science, Nanjing University, Nanjing 210093, China}
 \and {Center for Astrophysics, Guangzhou University, Guangzhou 510006, China}
 \and {School of Materials Science and Engineering, Baise University, Baise 533000, China}
}

\abstract
{Inflows at the dusty torus and smaller scales is crucial to investigate the process of supermassive black hole accretion. However, only few cases of inflowing gas at small scales have been reported through redshifted broad absorption lines so far. Here we report 9 redshifted narrow absorption lines (NALs) of $\rm Mg^+$ ions with inflowing speeds of 1071 --- 1979 \kms, which are likely along the directions close to the axes of accretion disks. The  quasars showing inflowing \MgII\ NALs have on average slightly smaller Eddington ratios when compared to the sources with outflow \MgII\ NALs. The upper limits of locations of the detected NALs are at parsec scale, around the distances of dusty tori to central SMBHs. The one possible origin of these infalling NALs is from dusty tori. Of course, these infalling NALs can also be naturally explained by chaotic cold accretion resulted from the nonlinear interaction of active galactic nucleus (AGN) jets with the interstellar medium, and these cold gaseous blobs may originally precipitate in metal-rich trailing outflows uplifted by AGN jet ejecta. The infalling NALs may thus provide direct evidence for cold gas precipitation and accretion in AGN feedback processes, and provide the direct evidence of inflowing gas along the directions close to quasar jets and at parsec scale.  It does not matter whether these infalling NALs are from the dusty tori or the interaction of AGN jets with the ISM, the infalling NALs cannot provide sufficient fuels to power the quasars.}
\keywords{galaxy physics --- galaxies: active --- quasars: absorption lines --- galaxies: infall} 
\maketitle

\section{Introduction}
Quasars are one of the brightest astronomical objects powered by accretion disks surrounding supermassive black holes (SMBHs). Numerous evidences \citep[e.g.,][]{2000ApJ...539L...9F,2009MNRAS.397.1705G,2010MNRAS.401.2531A,2014ARA&A..52..415M} have revealed that SMBHs and their host galaxies co-evolve \citep[e.g.,][]{2013ARA&A..51..511K,2014ARA&A..52..589H}. Feed and feedback are fundamental processes resulting in the co-evolution of SMBHs and galaxies. In the forms of jets, outflows, winds, and radiations, quasar feedback might heat and blow off the gas from central region of the host galaxy \cite[e.g.,][]{2000ApJ...545...63E,2017ApJ...837..149G}. This process likely compress the star formation within host galaxy, regulate the galaxy growth to be become overmassive, and impact on the surrounding circumgalactic medium (CGM) and intergalactic medium (IGM).

The trigger of nuclear activity requires gas supply, from a fraction to dozens of solar masses per year \citep[e.g.,][]{2019NatAs...3...48S}, to feed the Active Galactic Nuclei (AGN). Although galaxy interactions and mergers can efficiently trigger and feed the quasars \citep[e.g.,][]{2007ApJ...655..718S,2012ApJ...758L..39T,2014A&A...569A..37M,2015A&A...576A..32G,2017ApJ...837..149G,2019NatAs...3...48S}, some fundamental questions are still unanswered. How the accretion disks are supplied with external gas? Whether dose all the inflowing gas reach at accretion disk? Where is the inflowing gas originated from? The answers are benefit to comprehend the feed and feedback mechanisms and processes of quasars. Therefore, the observational evidence of inflows at the dusty torus and smaller scales (less than several tens of parsec) is crucial to study the process of SMBH accretion. Cold clouds in outflows, winds, or surrounding CGMs will condense and fall back toward central regions, providing material for broad emission line regions and central black holes \cite[e.g.,][]{2013MNRAS.432.3401G,2014ApJ...789..153L,2017ApJ...847...56E,2017ApJ...837..149G}. Many previous works have suggested inflowing clouds in/around the broad emission line region of quasars \cite[e.g.,][]{2008ApJ...687...78H,2009ApJ...707L..82F,2013ApJ...769...30G,2016Ap&SS.361...67G,2017ApJ...849..146G}.  However, up to now, only a few cases \citep[e.g.,][]{2013MNRAS.434..222H,2017ApJ...843L..14S,2017ApJ...839..101Z,2019Natur.573...83Z} of inflowing gas at these scales have been reported through redshifted broad absorption lines (BALs), where the BALs are defined as the absorption features hosting continual absorption features with line widths $>2000$ \kms\ at depths $>10\%$ below the continuum \citep[][]{1979ApJ...234...33W}, and are likely close to the equatorial plane of accretion disk. In addition, it is vacant up to now that the unambiguously observational evidence of inflowing gas is detected through metal narrow absorption line systems (NALs) in quasar spectra, where the NALs are the absorption features with line widths less than a few \kms, and are likely along the directions close to the axes of accretion disks \citep[][]{2012ASPC..460...47H}. In theory, clouds with enough column density of gas, which intercept quasar sightlines, could leave absorption features in quasar spectra. If the absorbing gas cloud falls towards the quasar central region (accretion disk), we can observe inflowing (redshifted) absorption features. This vacantness of the redshifted metal NALs can be ascribed to several factors. The first one is the quasar systemic redshift, which should be robust and accurate. If the quasar redshift is determined from broad emission lines, especially from the highly ionized \CIV, there may be a large uncertainty in the quasar redshift \citep[e.g.,][]{2011ApJS..194...45S}, which often leads to misidentification of inflowing NALs. The second factor is that a NALs should be identified through at least 3 robust narrow absorption lines at the same redshift, which significantly reduces the sample size of NALs. If a NALs is only confirmed through the \NVab, \CIVab, or \MgIIab\ resonance doublet, the probability of false absorption line systems cannot be ignored. The third factor is that the radiation from quasar central regions is strong along the directions close to the axes of accretion disks. Most of the inflowing gas clouds, which are expected to produce NALs and along the directions close to the axes of accretion disks, are likely transformed into outflows by the strong quasar radiation when they reach hundreds of gravitational radii from SMBHs \citep[][]{2017ApJ...837..149G}. This would significantly reduce the incidence of inflowing NALs. Therefore, the inflowing NALs with high significant level are very difficult to be identified.

Here we report a sample of 9 redshifted narrow absorption line systems, whose upper limits of locations are around the dusty tori. The inflowing NALs possibly provide direct evidence for cold gas precipitation and accretion in AGN feedback processes, and provide the first direct evidence of inflowing gas along the directions close to quasar jets and at parsec scale. In this paper, we adopt the $\rm \Lambda CDM$ cosmology with $\rm \Omega_M=0.3$, $\rm \Omega_\Lambda=0.7$, and $\rm H_0=70~km~s^{-1}~Mpc^{-1}$.

\section{Data sample and spectral analysis}\label{sect:data}
We aim to study quasar inflows through narrow absorption lines imprinted in quasar spectra. Therefore, we firstly select \MgII\ NALs with $\upsilon_r < 0$ from the largest catalog of quasar \MgII\ associated absorption lines \citep[][]{2018ApJS..235...11C}, which were detected from about $10^5$ quasar spectra of the Sloan Digital Sky Survey \citep[SDSS,][]{2017A&A...597A..79P}. Here
\begin{equation}\label{eq:vr}
\upsilon_r = c\times\frac{(1 + z_{em})^2 - (1 + z_{abs})^2}{(1 + z_{em})^2 + (1 + z_{abs})^2},
\end{equation}
where the $z_{\rm abs}$ is the absorption line redshift, the $z_{\rm em}$ is the quasar systemic redshift, and the $c$ is the speed of light. Generally speaking, an absorber with $\upsilon_r < 0$ means that it is likely falling into quasar center region. The $z_{\rm abs}$, which is measured from the narrow absorption line, is accurate. While, due to some mechanisms, such as quasar outflow, the $z_{\rm em}$, which is determined from quasar emission lines, is usually less than the true value. Therefore, it requires accurate quasar systemic reshift to define an inflow absorber by the velocity offset between \zem\ and \zabs. The quasar systemic redshift determined from narrow emission lines, such as \OII, is more accurate than that estimated from broad emission lines. Therefore, we further limit absorber sample to the \MgII\ NALs, which were detected in the quasar spectra shown obvious \OII\ emission features. Meanwhile, we determine all the quasar systemic redshifts by fitting the \OII\ emission features.

\citet{2018ApJS..235...11C} identified a \MgII\ absorption line system only by the \MgIIab\ doublet. Only two absorption lines detected at the same redshift may lead to some false \MgII\ absorption line systems. In order to enhance the reliability of \MgII\ absorption line systems, we require that: (1) except for the \MgIIab\ doublet, one absorption system has at least two absorption lines at other rest-frame wavelengths that are detected at the same redshift, such as the series of absorption lines of \FeII, the \MgI; and (2) the absorption line strengths with $W_r^{\lambda2796}>5\sigma_{W_r^{\lambda2796}}$ and $W_r^{\lambda2803}>3\sigma_{W_r^{\lambda2803}}$.

In term of the criteria mentioned above, we find that there are 9 \MgII\ NALs with $\upsilon_r < -1000$ \kms\ in the absorption line catalog of \citet{2018ApJS..235...11C}. Redshifted \MgII\ NALs can be originated from (1) measured errors of both the absorption line and emission line redshifts, due to the dispersions of absorption and emission lines; (2) \OII\ emission line redshifts that cannot well represent the quasar systemical redshifts, since gas flows within emission line region may lead to an asymmetrical line profile; (3) the \MgII\ NALs formed within the external galaxies that are randomly moving in a cluster with large mass. The absorption features of all the 9 \MgIIab\ doublets are very significant and have good line profiles. In addition, two lines of all the \MgIIab\ doublets are clearly resolved with each other, and their line widths ($\sigma_{\rm abs}$) are significantly less than 200 \kms. Therefore, the absorption redshifts ($z_{\rm abs}$) of the \MgII\ NALs with $\upsilon_r < -1000$ \kms\ determined from the \MgIIab\ doublets are accurate. We determine the quasar systemic redshifts through \OIIwave\ narrow emission lines, which have line widths ($\sigma_{\rm eml}$) about 300 \kms. Furthermore, both of the \MgII\ and \OII\ emission lines have symmetrical profiles for all the quasars with redshifted \MgII\ NALs. In other words, we don't observe obvious outflow features in the broad \MgII\ and narrow \OII\ emissions. Therefore, the quasar redshifts ($z_{\rm em}$) determined from \OII\ are robust. Relative to the \OII\ emission lines, the velocity offsets of the 9 redshifted \MgII\ NALs are $\upsilon_{\rm r}=1071$ --- 1979 \kms, which are significantly larger than the uncertainty ($\sigma_{\rm drift}$) contributed from line dispersions of the \MgII\ NALs and \OII\ emission lines ($\sigma_{\rm drift} = \sqrt{\sigma_{\rm abs}^2 + \sigma_{\rm eml}^2}=\sqrt{200^2 + 300^2}=360$ \kms). Therefore, the large velocity offsets ($\upsilon_r < -1000$ \kms) of the 9 redshifted \MgII\ NALs are less likely originated from line dispersions.

\begin{figure}
\centering
\includegraphics[width=0.43\textwidth]{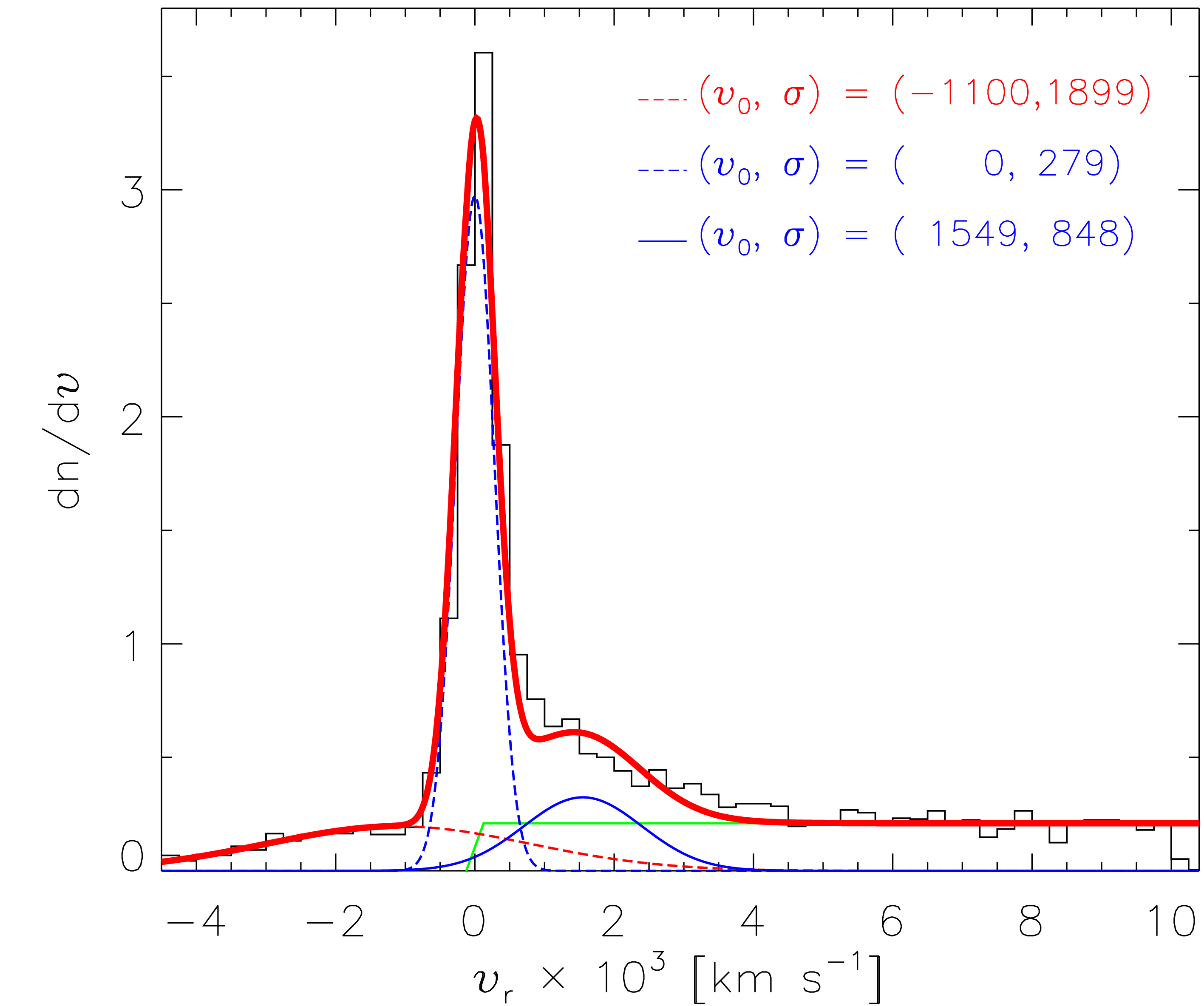}\\
\caption{Relative velocity distribution of 6349 \MgII\ NALs with $z_{em}<1.07$. The data are directly taken from the $17,316$ \MgII\ NALs of \citet{2018ApJS..235...11C}. The red dashed, blue dashed, and blue solid curves indicate the Gaussian function fits, which correspond to inflow, environment, and outflow \MgII\ NALs, respectively. The green solid line indicates the mean count at $\upsilon_r>6000$ \kms, which corresponds to intervening \MgII\ NALs. The red solid curve is the sum of all the color lines. The values shown in the top right corner are the Gaussian function fitting centers ($\upsilon_0$) and dispersions ($\sigma$).}
\label{fig:dopp}
\end{figure}

It is well accepted that the quasar resides within a host galaxy, which is surrounded by circumgalactic medium (CGM). In addition, the quasar host galaxy would be a member galaxy of a galaxy cluster. Absorption lines would be imprinted in the quasar spectra, when the quasar center emission passes through its surrounding gas clouds located within an outflow, inflow, host galaxy, CGM, and intergalactic medium (IGM) of the galaxy cluster before it reaches the observer. Therefore, a clustering distribution of absorption systems is often detected around the quasar emission redshift  relative to the cosmological absorptions \citep[e.g.,][]{2008MNRAS.386.2055N,2008MNRAS.388..227W,2015ApJS..221...32C,2018ApJS..235...11C}. We reanalyze the velocity distribution of \MgII\ NALs included in \citep{2018ApJS..235...11C}, after cutting their quasars with $z_{\rm em}<1.07$ so that the quasar redshifts can be better determined from emission lines \citep[e.g.,][]{2018ApJS..234...16C}. This cut reduces the numbers of \MgII\ NALs from $17,316$ of \citet{2018ApJS..235...11C} to 6349 of this paper. The velocity distribution of these 6349 \MgII\ NALs is shown in Figure \ref{fig:dopp}, which clearly shows that there is an significantly excess around $\upsilon_r\approx0$. This suggests that the \MgII\ absorptions would be clustered around quasars. In addition, Figure \ref{fig:dopp} also implies that velocity distribution of \MgII\ absorbers is complex, and the absorber sample should include absorptions that originated in the quasar's outflow, inflow, surrounding environment, and foreground intervening galaxies. The intervening absorbers should have a uniform random velocity distribution at $\upsilon_r>0$. The environmental absorbers should have a normal velocity distribution at $\upsilon_r=0$, which is destroyed by the outflow absorbers at blue wing ($\upsilon_r>0$). The extended red wing of Figure \ref{fig:dopp} ($\upsilon_r<0$) is likely originated from inflow absorbers. Accounting for the multiple components of the velocity distribution, we invoke three Gaussian functions to describe velocity distributions of the outflow (Gaussian center $\upsilon_r>0$), environment (Gaussian center $\upsilon_r=0$), and inflow (Gaussian center $\upsilon_r<0$) absorbers, and invoked a linear function with slope $\alpha=0$ to characterize the velocity distribution of the intervening absorbers. The results are exhibited with color lines in Figure \ref{fig:dopp}, which clearly shown that all the inflow, environmental, and outflow absorbers are mixed in the velocity range of $\upsilon_r<0$. In addition, both the environment and outflow \MgII\ NALs are limited within $\upsilon_r>\upsilon_0 -
 3\sigma>-1000$ \kms, where the $\upsilon_0$ and $\sigma$ are the Gaussian function fitting centers and dispersions, respectively. Therefore, the ``$\upsilon_r<-1000$ \kms'' is a good velocity cut to select the true inflow absorbers, which can well eliminate the contamination from the environment/outflow absorptions. Thus, the 9 redshifted \MgII\ NALs with $\upsilon_r<-1000$ \kms\ are statistically unlikely to be formed within outflow or the external galaxies (environment) that are randomly moving in a cluster, but they are likely truly inflowing towards quasar central regions.

\begin{figure*}[htbp]
\centering
\includegraphics[width=12cm,height=2.5cm]{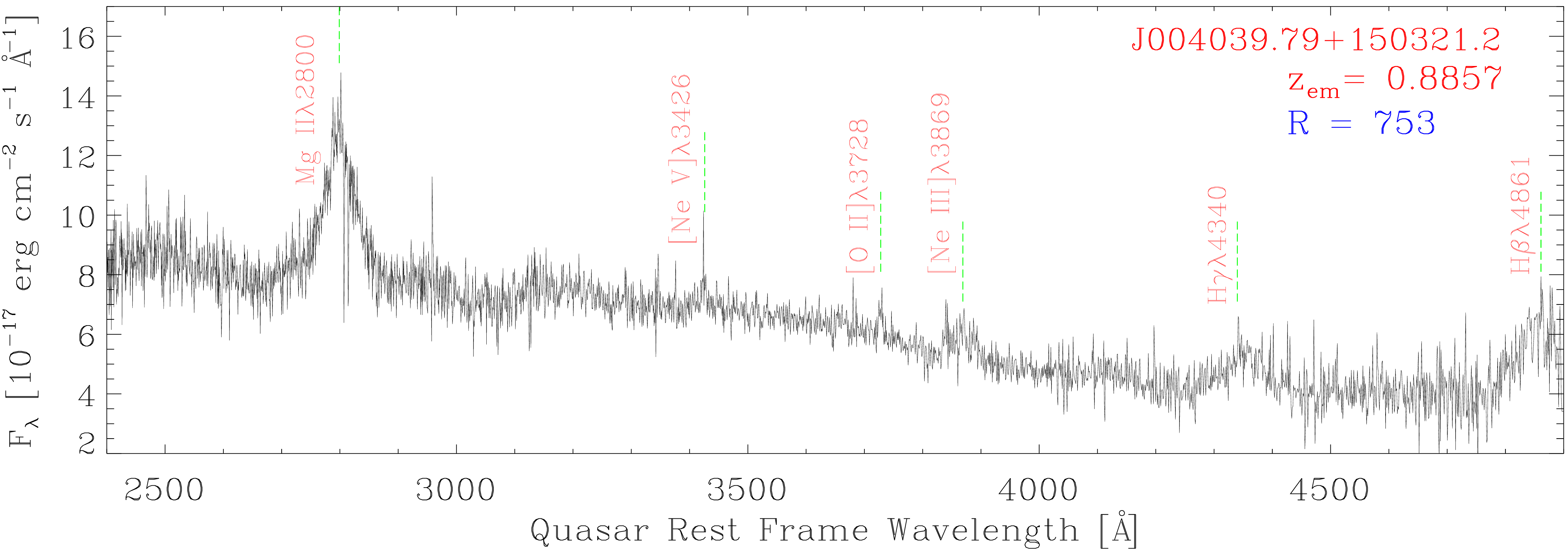}
\includegraphics[width=4.5cm,height=2.5cm]{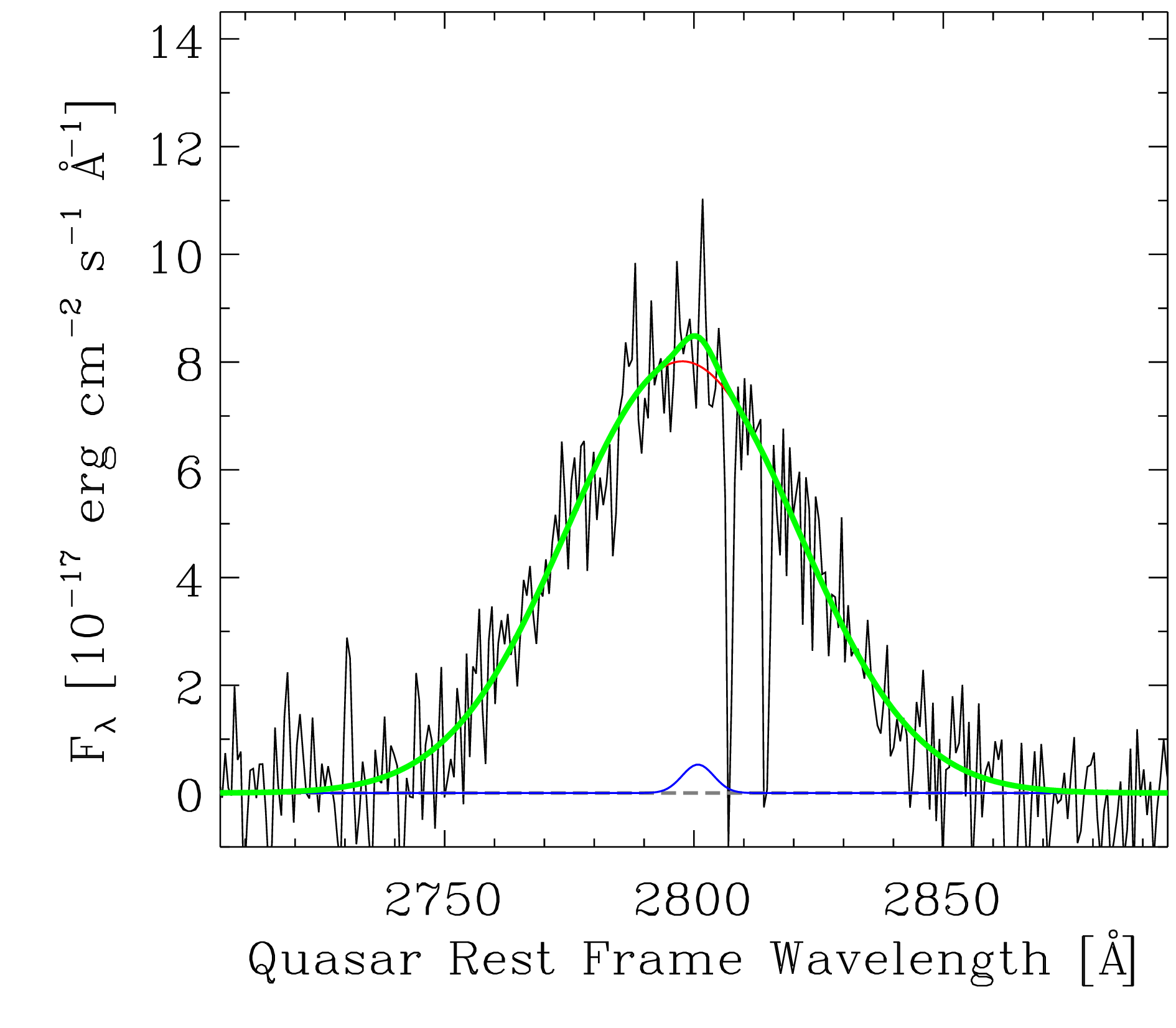}\\
\includegraphics[width=12cm,height=2.5cm]{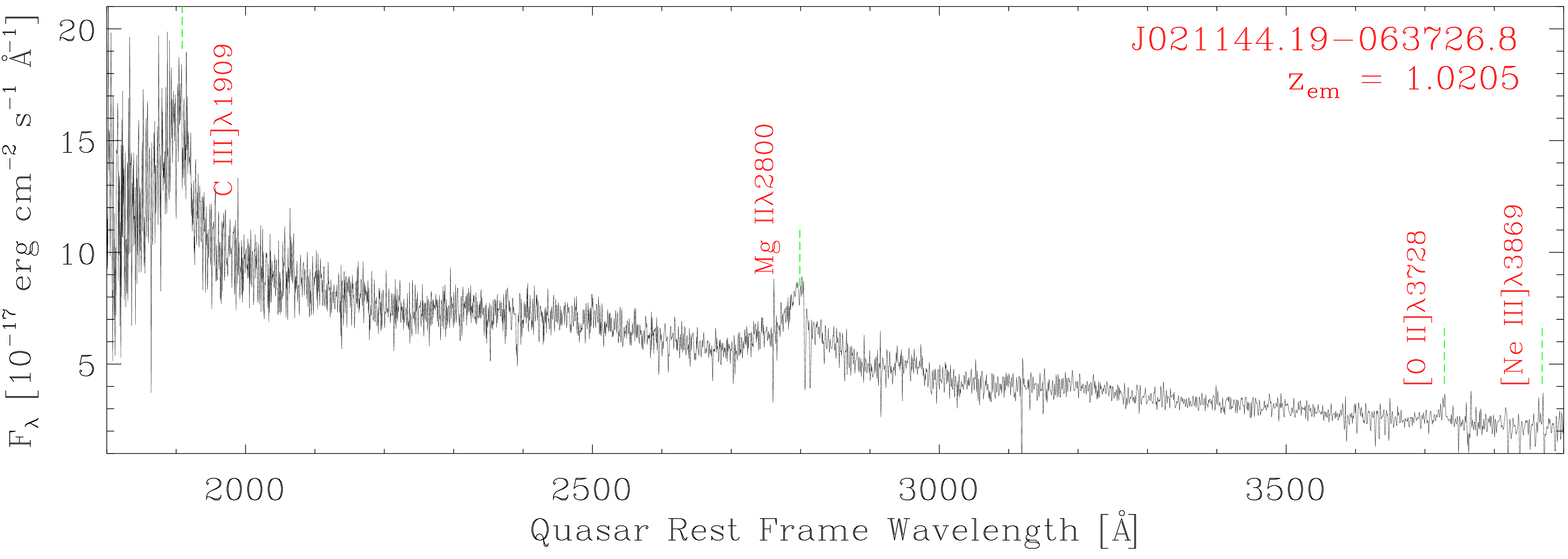}
\includegraphics[width=4.5cm,height=2.5cm]{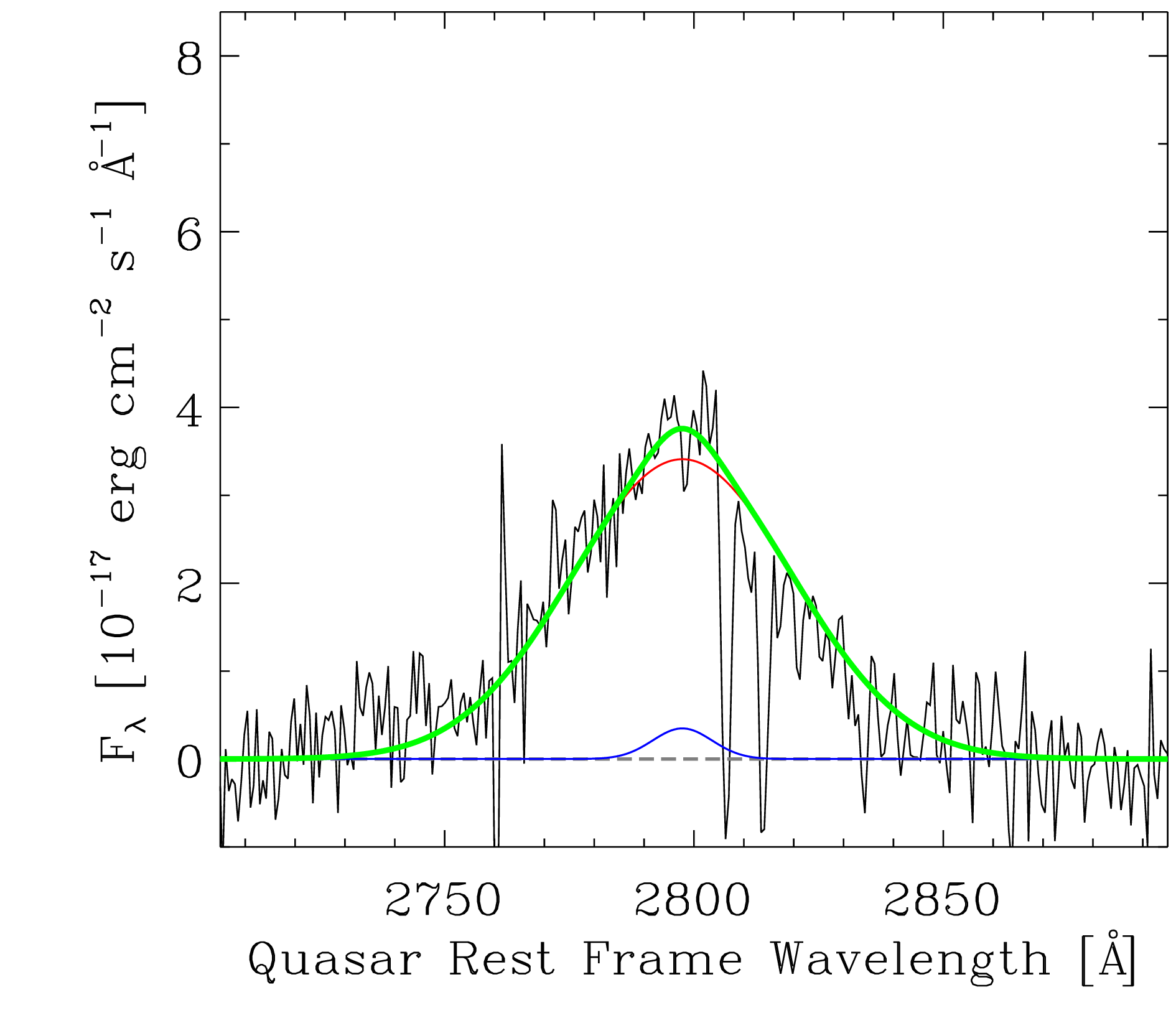}\\
\includegraphics[width=12cm,height=2.6cm]{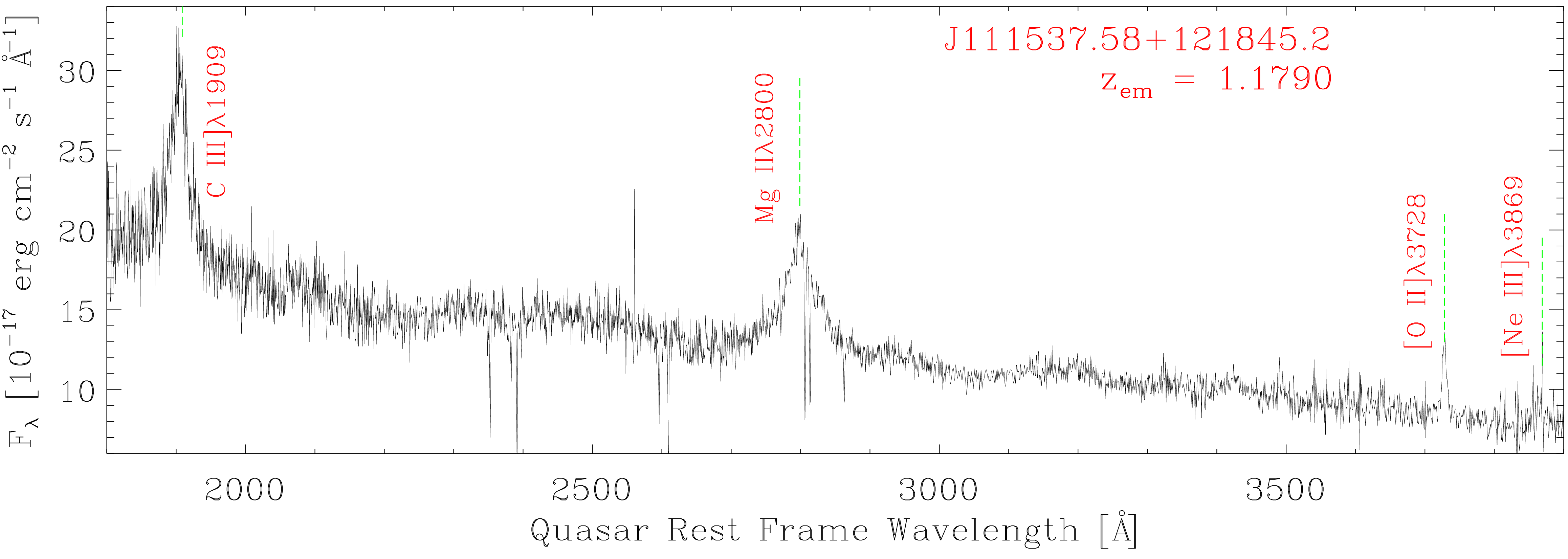}
\includegraphics[width=4.5cm,height=2.6cm]{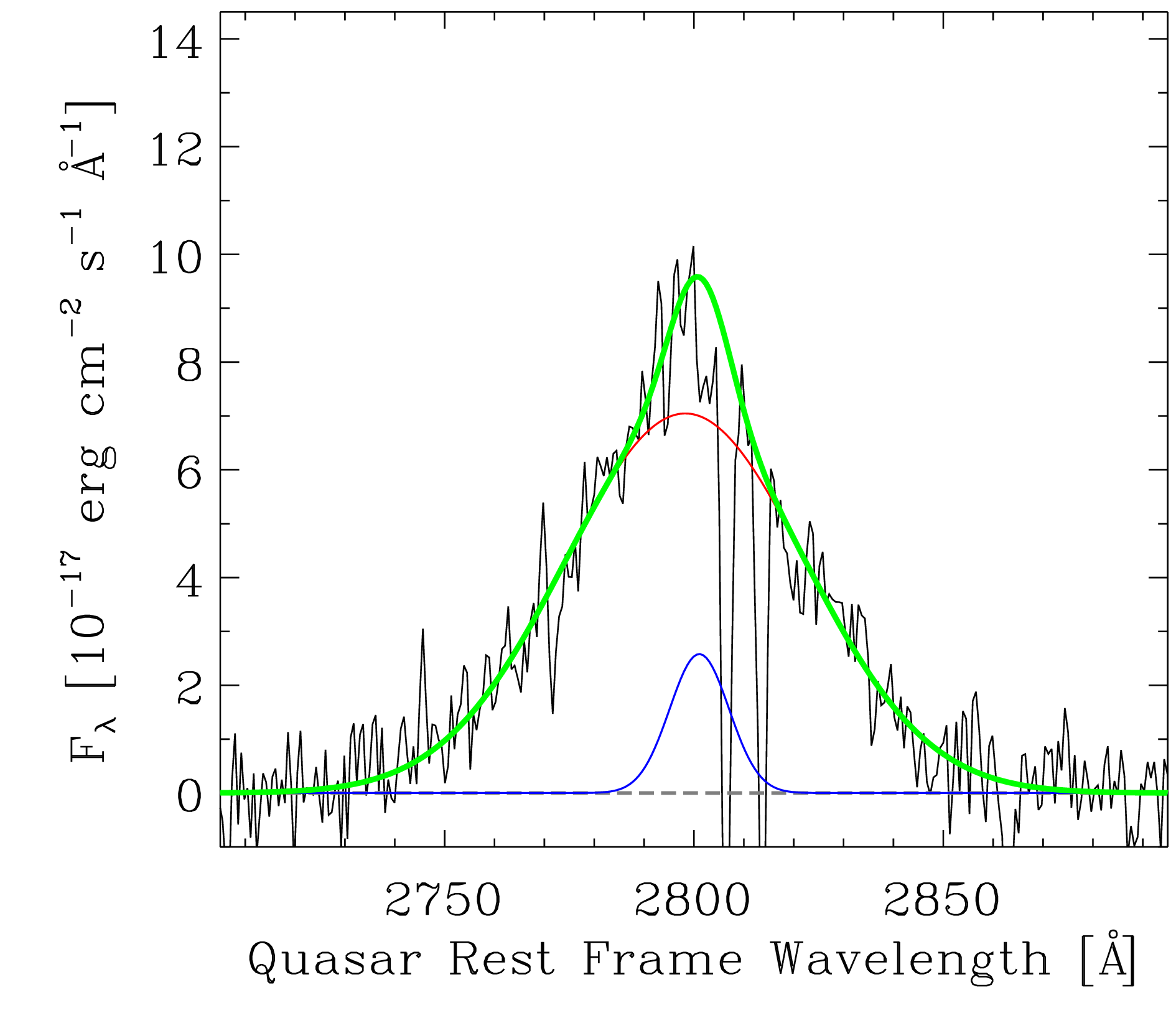}\\
\includegraphics[width=12cm,height=2.6cm]{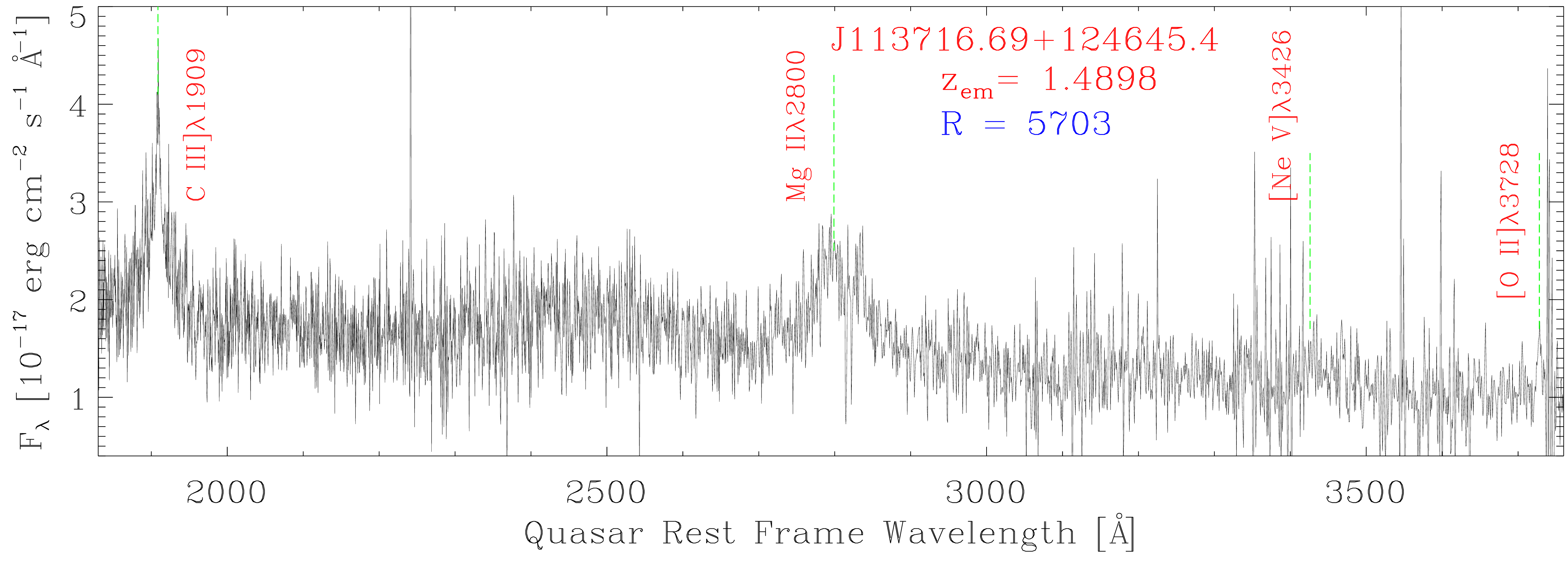}
\includegraphics[width=4.5cm,height=2.6cm]{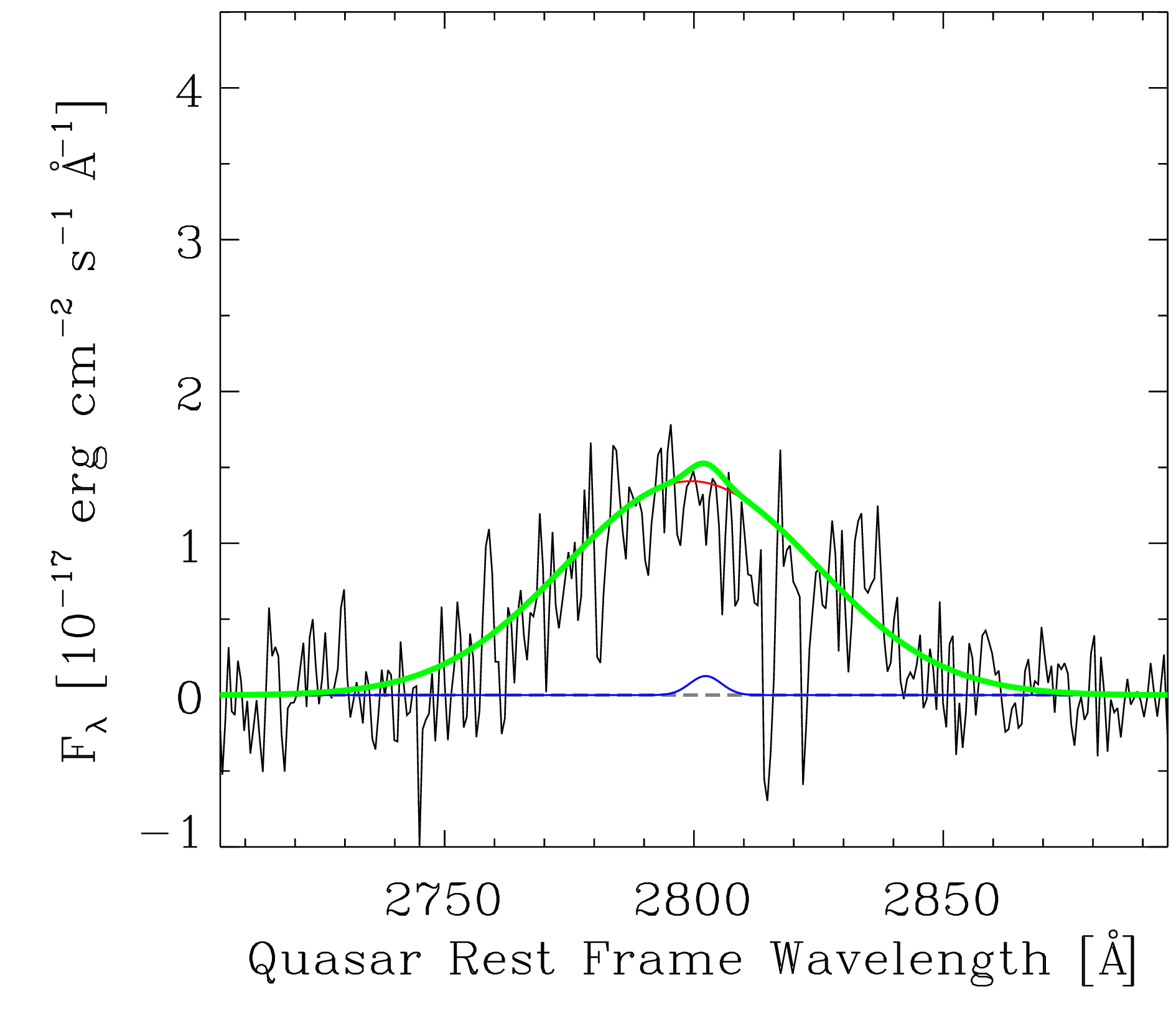}\\
\includegraphics[width=12cm,height=2.6cm]{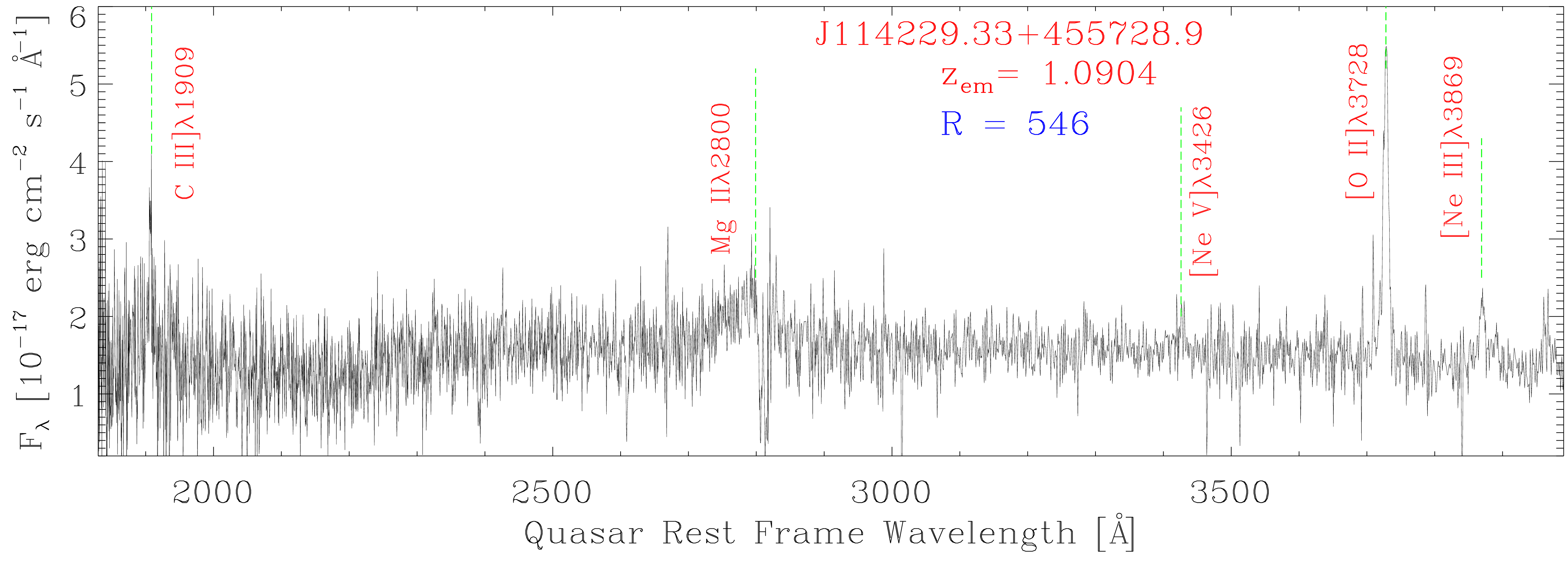}
\includegraphics[width=4.5cm,height=2.6cm]{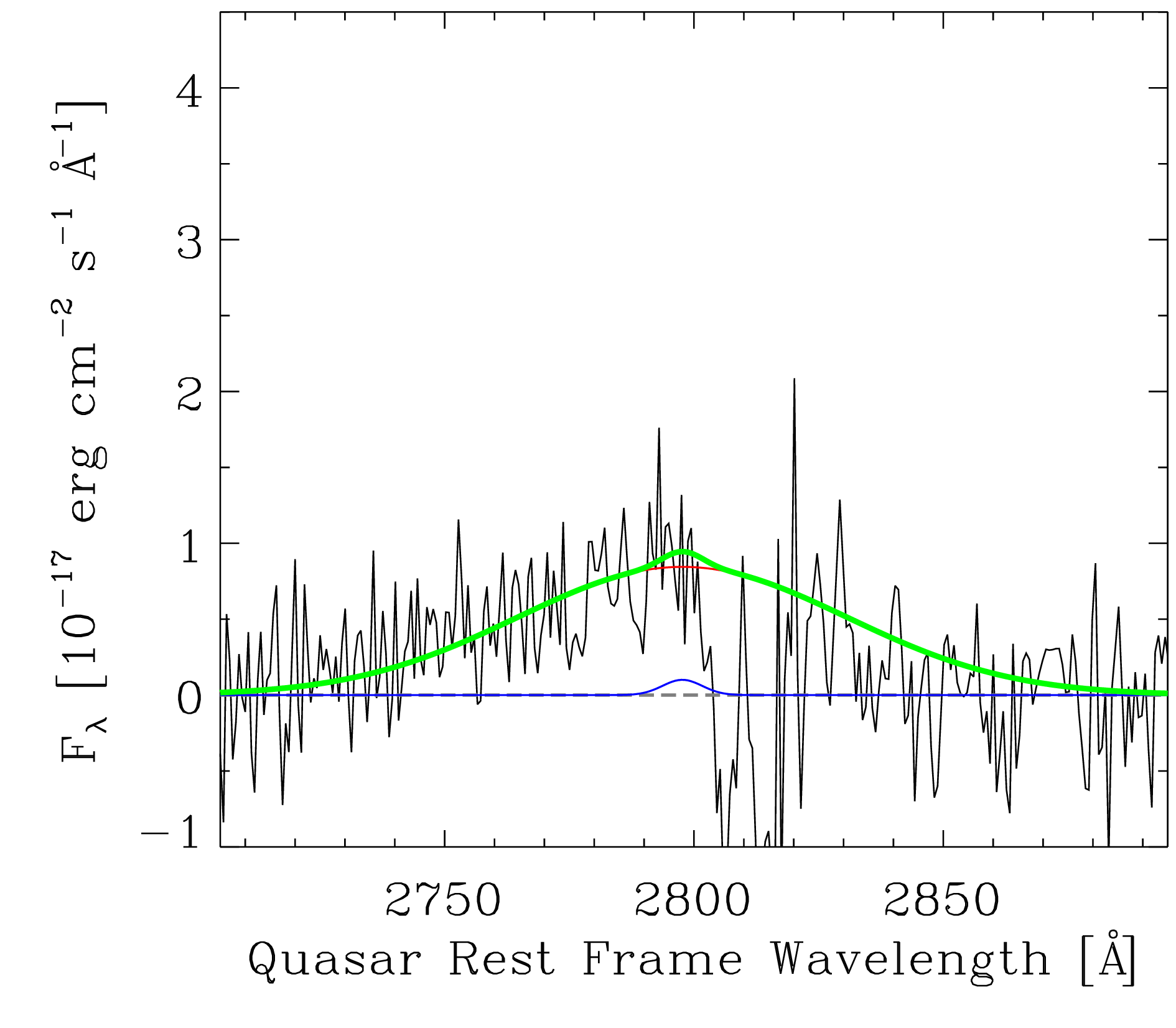}\\
\includegraphics[width=12cm,height=2.6cm]{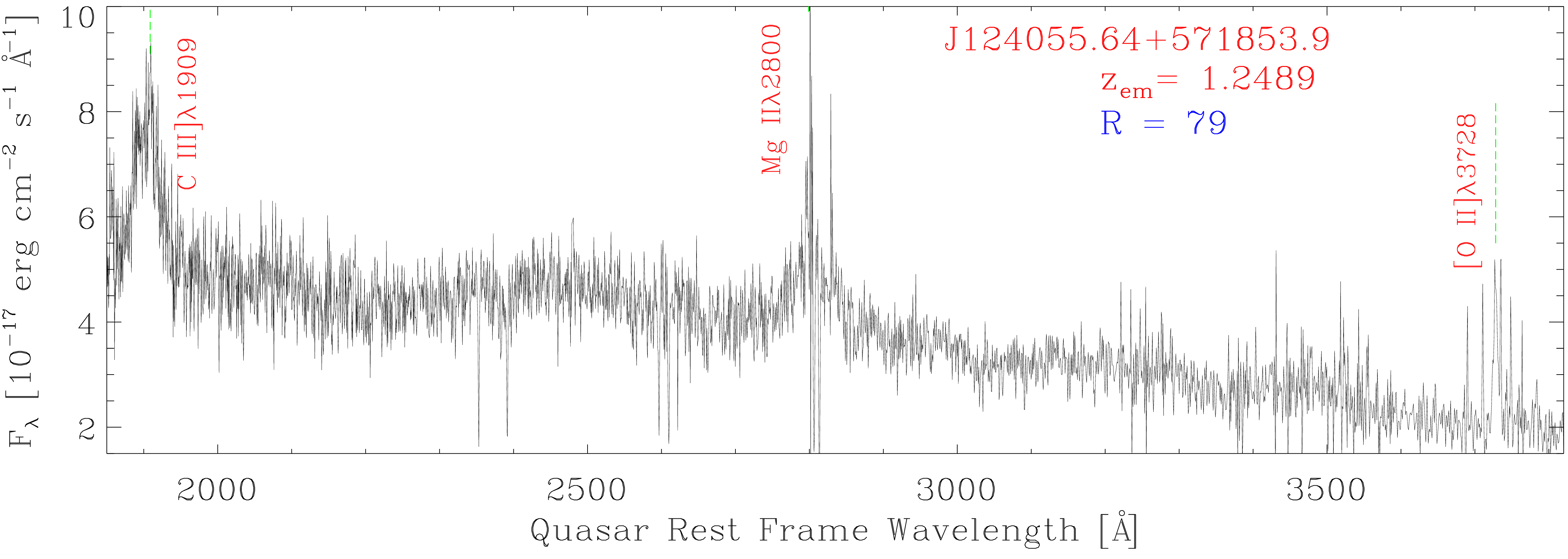}
\includegraphics[width=4.5cm,height=2.6cm]{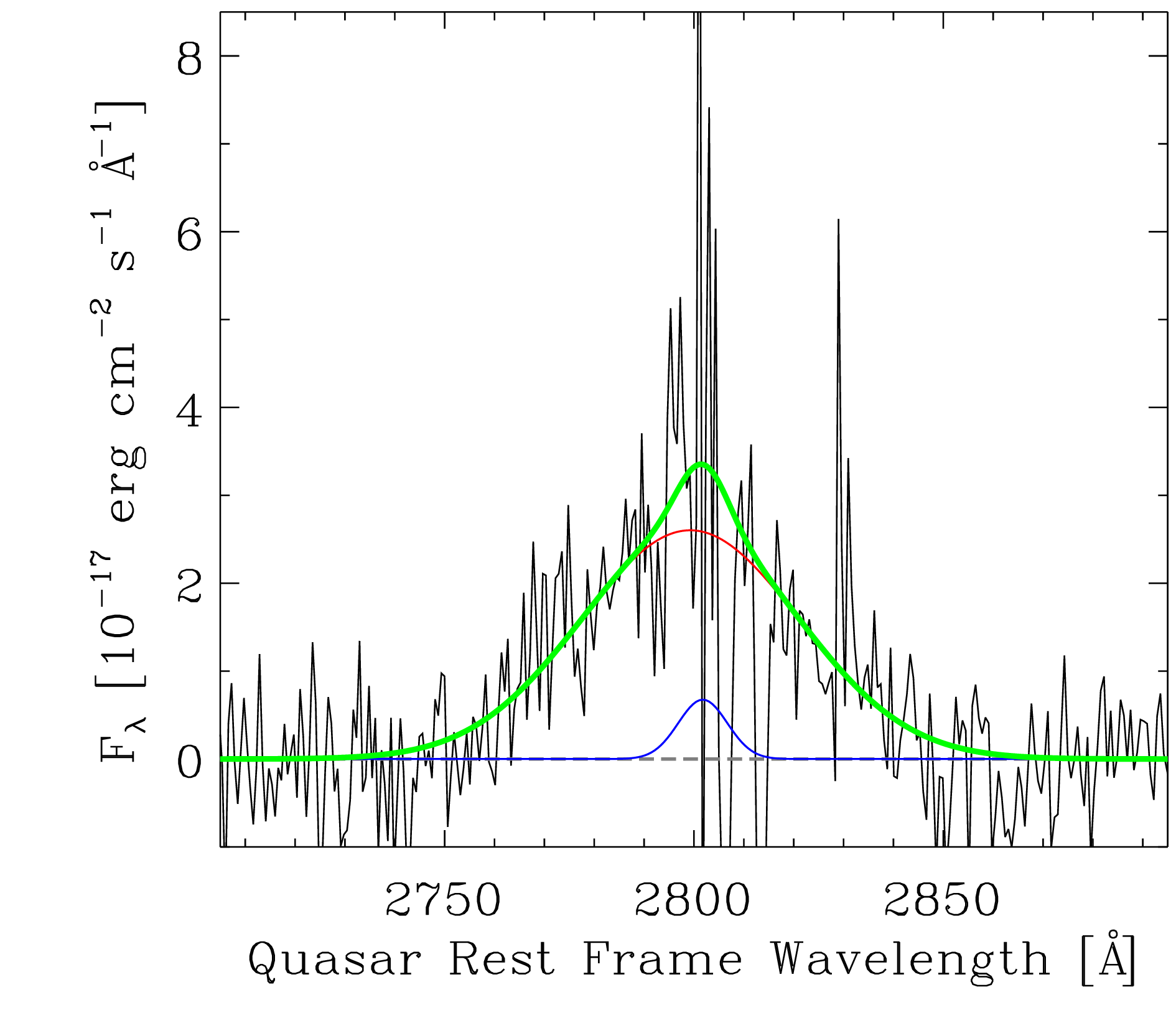}\\
\includegraphics[width=12cm,height=2.6cm]{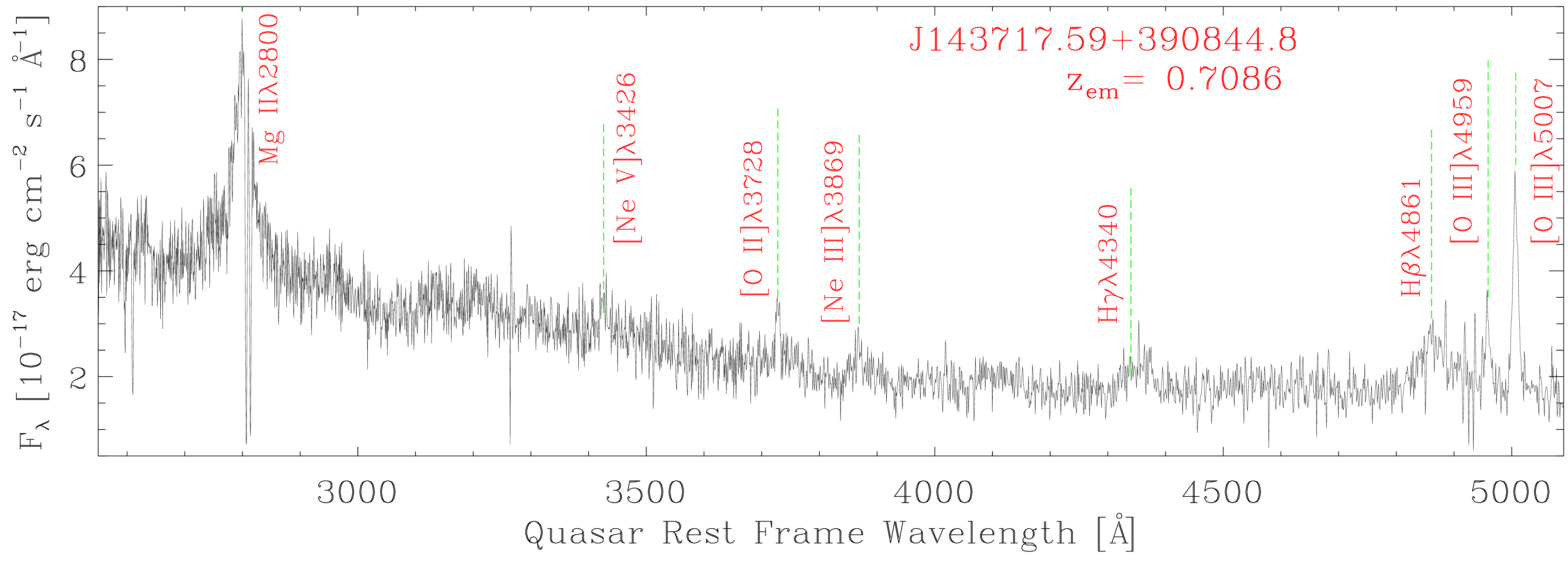}
\includegraphics[width=4.5cm,height=2.6cm]{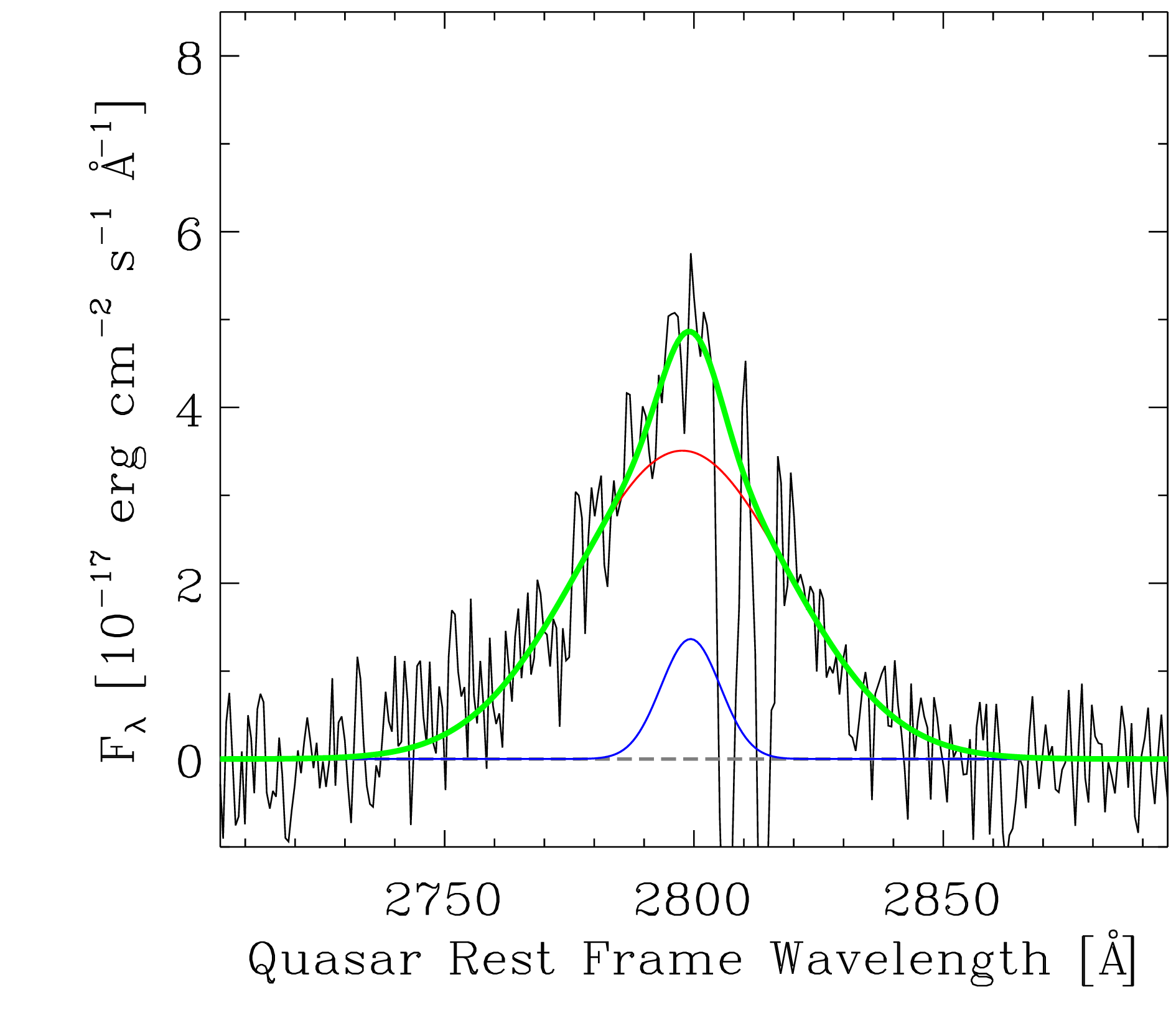}\\
\includegraphics[width=12cm,height=2.6cm]{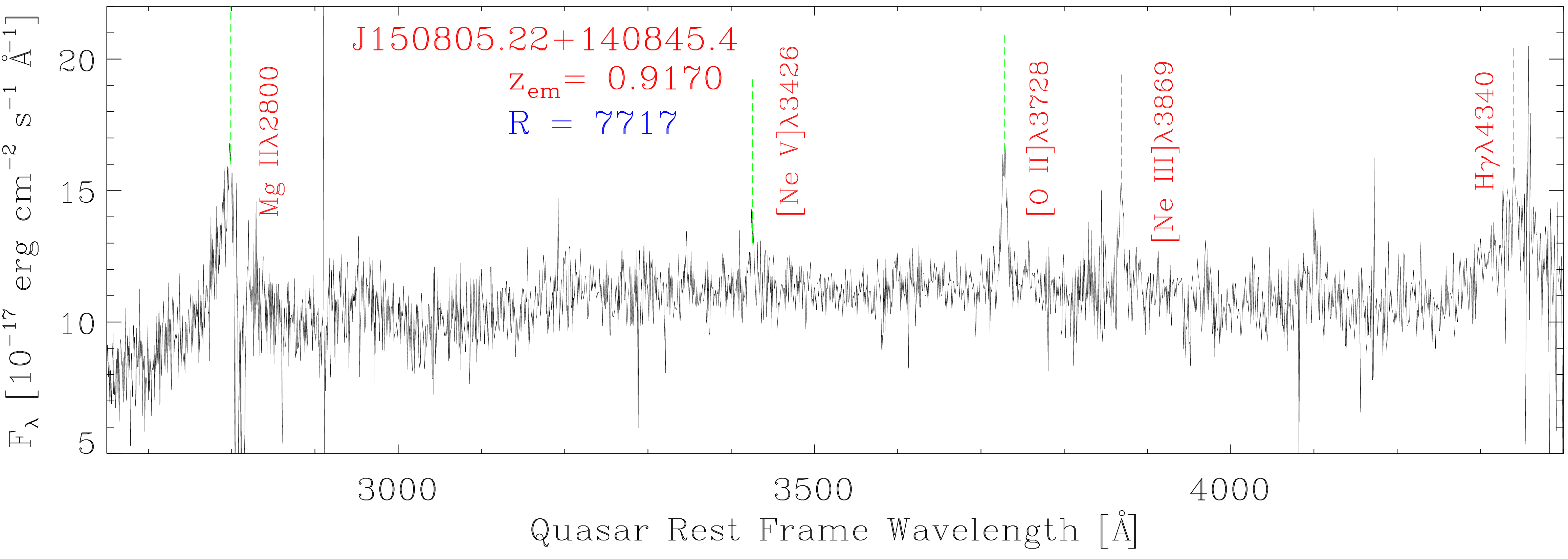}
\includegraphics[width=4.5cm,height=2.6cm]{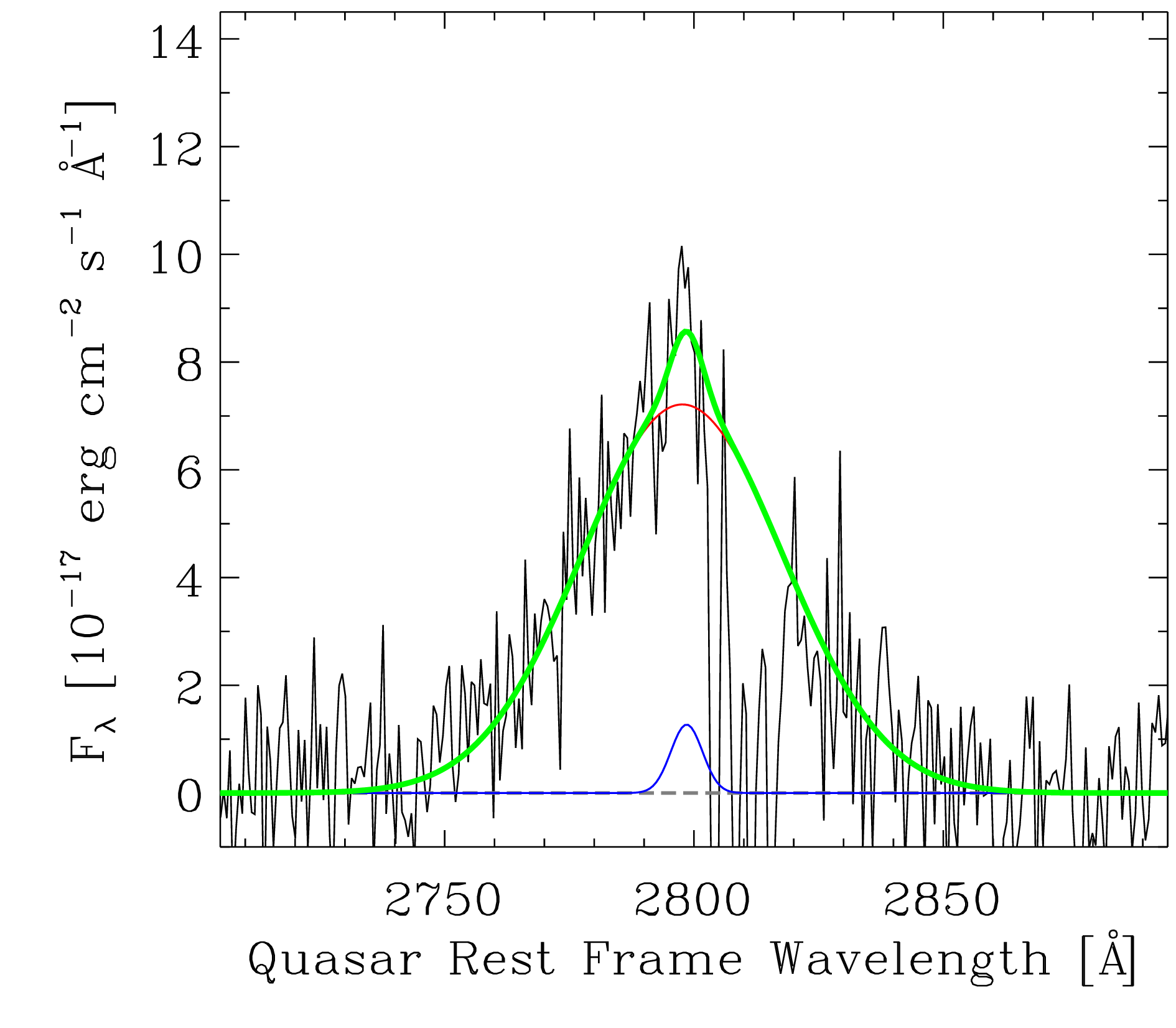}\\
\includegraphics[width=12cm,height=2.6cm]{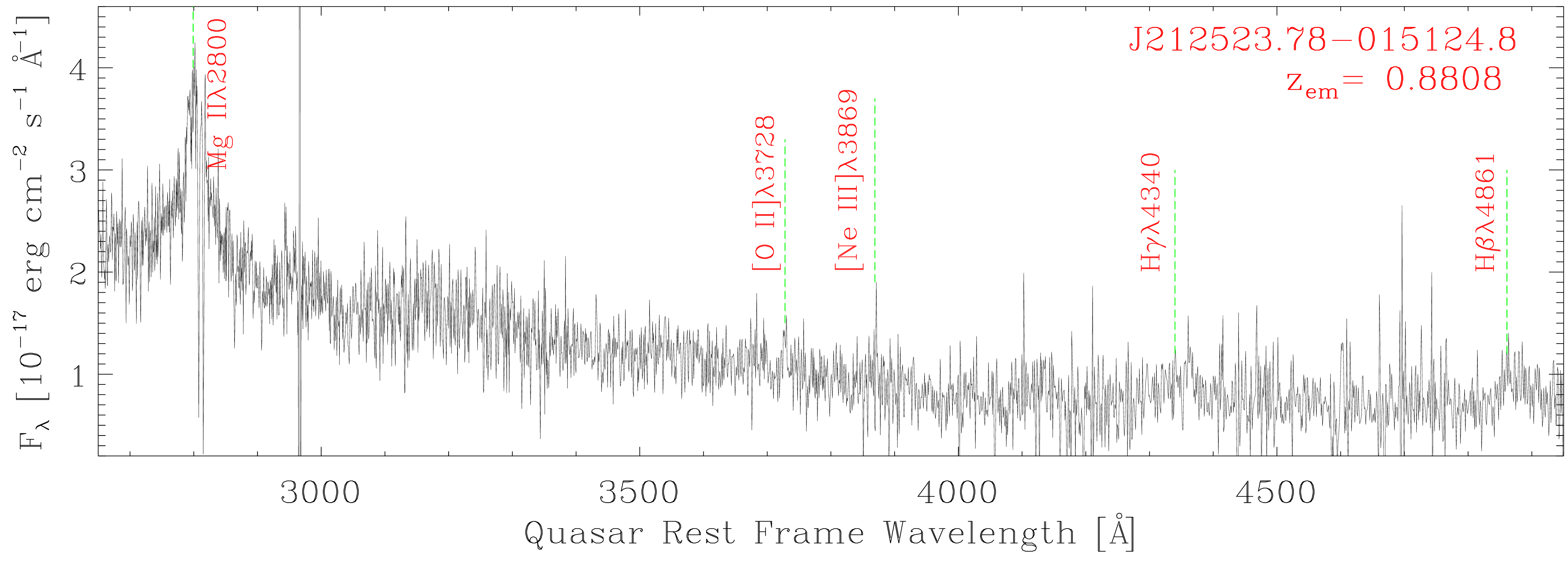}
\includegraphics[width=4.5cm,height=2.6cm]{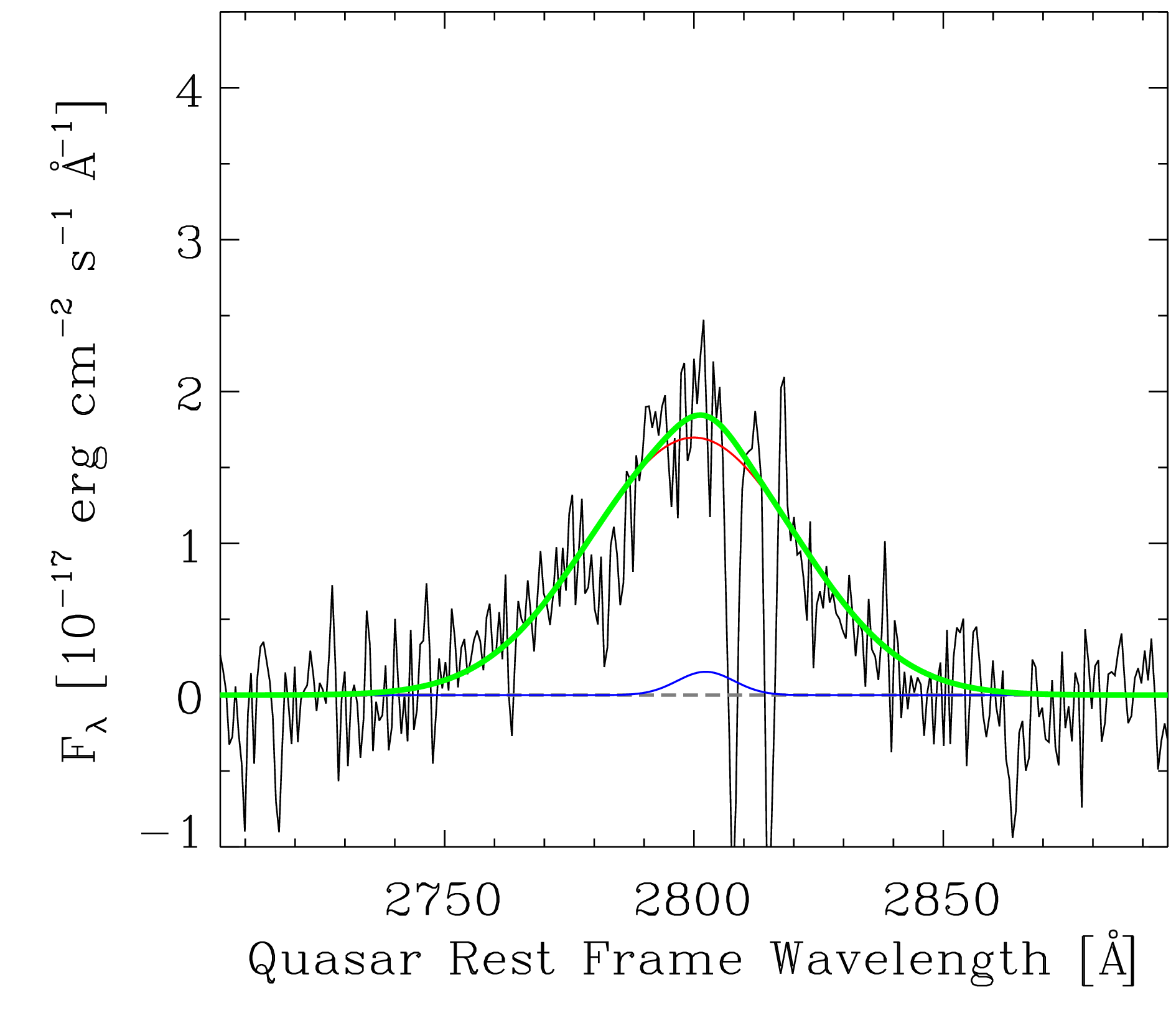}\\
\caption{The SDSS quasar spectra. Left panel: Green dash-lines indicate the positions of emission lines. The emission line redshifts $z_{\rm em}$ are determined from the \OII\ emission lines. The blue symbols are the radio loudness of quasars: $R=\frac{f_{6cm}}{f_{2500}}$. Rigth panel: The quasar spectra have been subtracted by the continuum+iron fits. The blue and red lines are the narrow and broad Gaussian function fits, respectively. The green lines are the sum of the blue and red lines.}
\label{fig:qso}
\end{figure*}

\begin{figure*} \centering
\includegraphics[width=12cm,height=2.6cm]{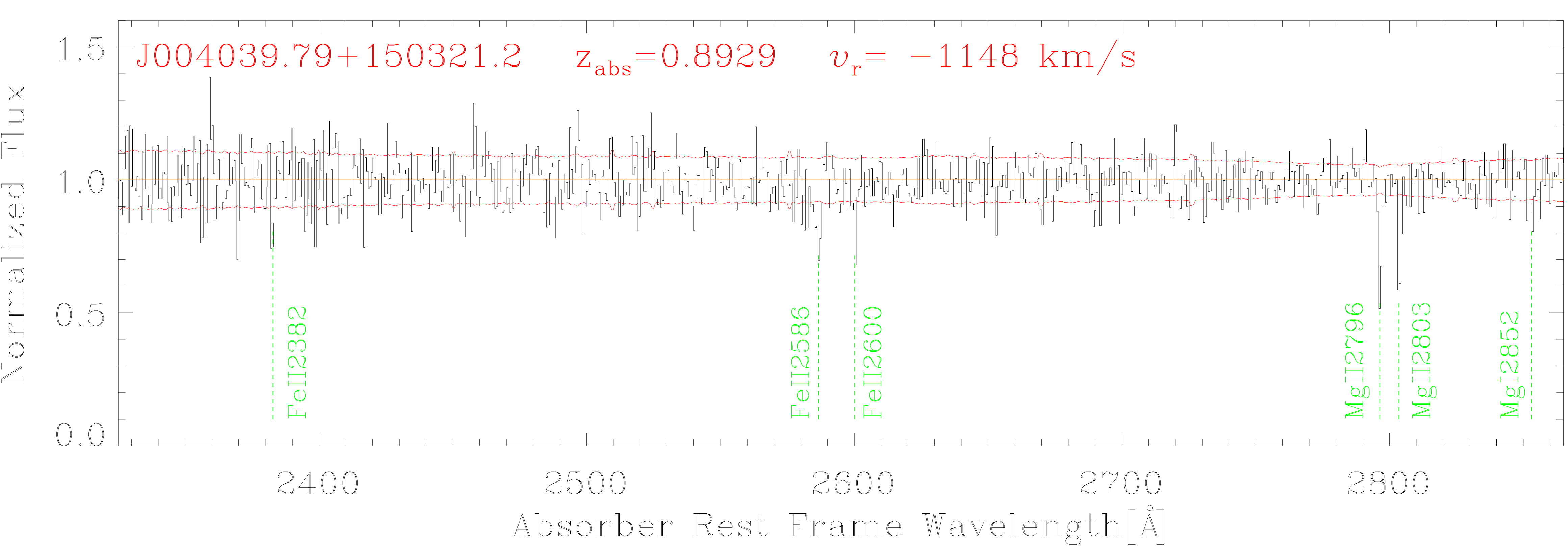}\\
\includegraphics[width=12cm,height=2.6cm]{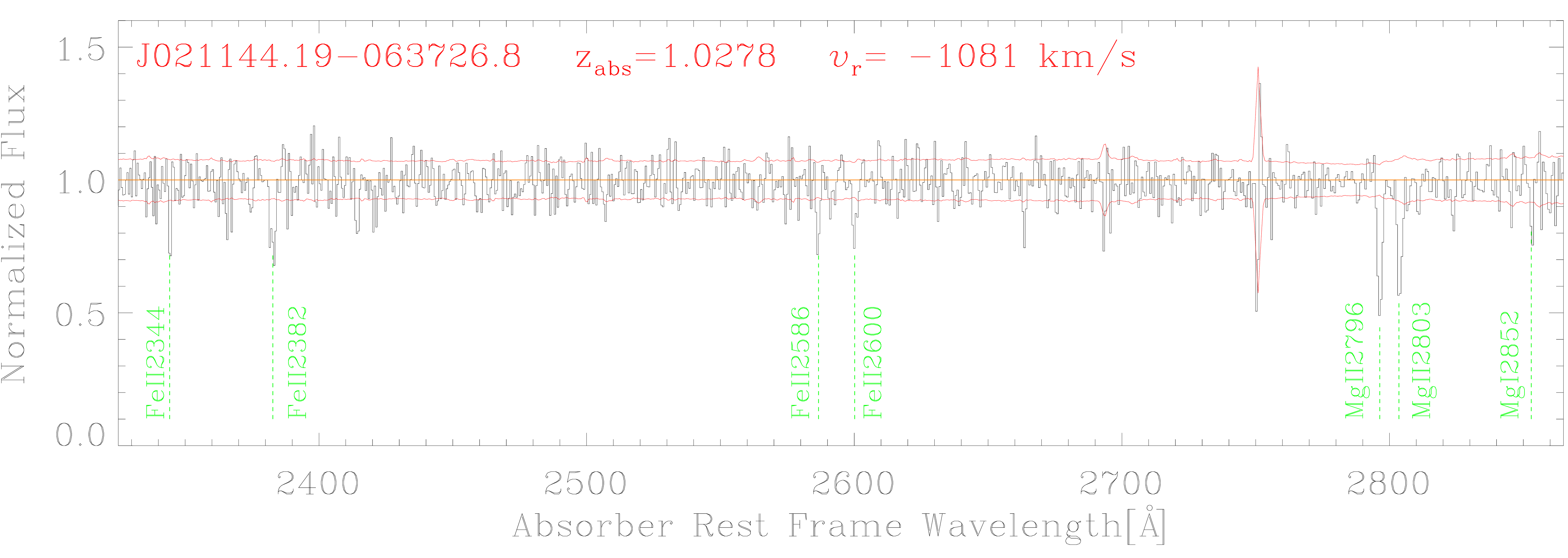}\\
\includegraphics[width=12cm,height=2.6cm]{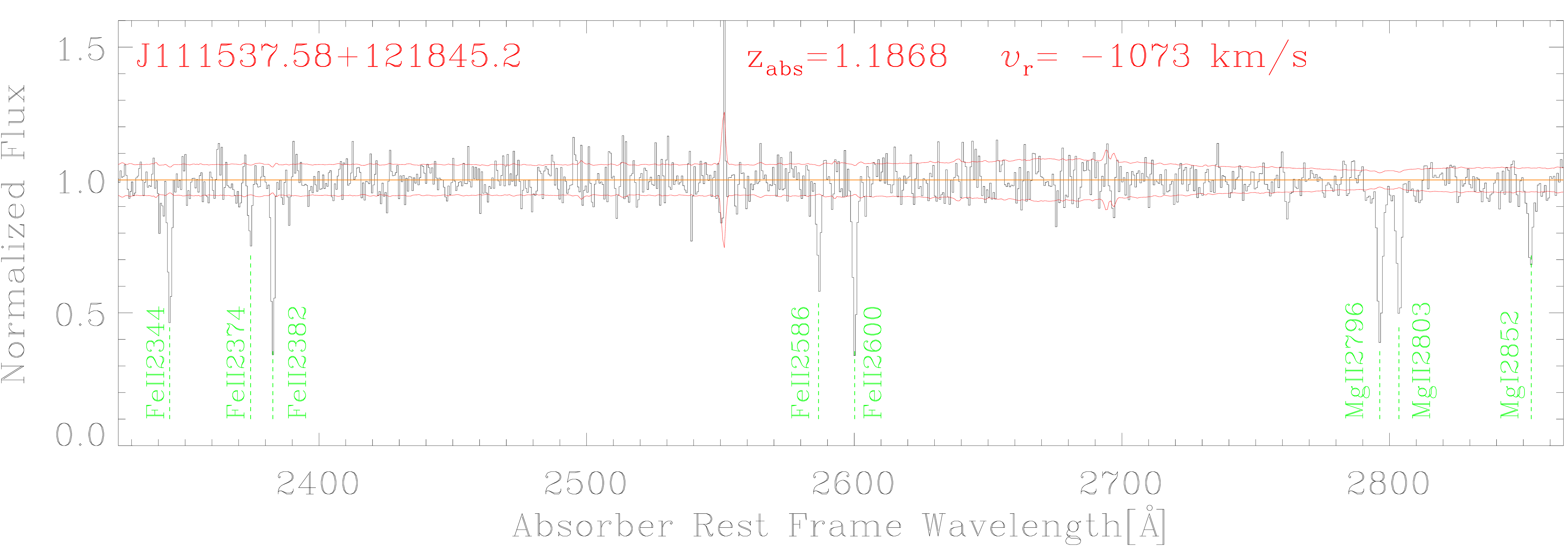}\\
\includegraphics[width=12cm,height=2.6cm]{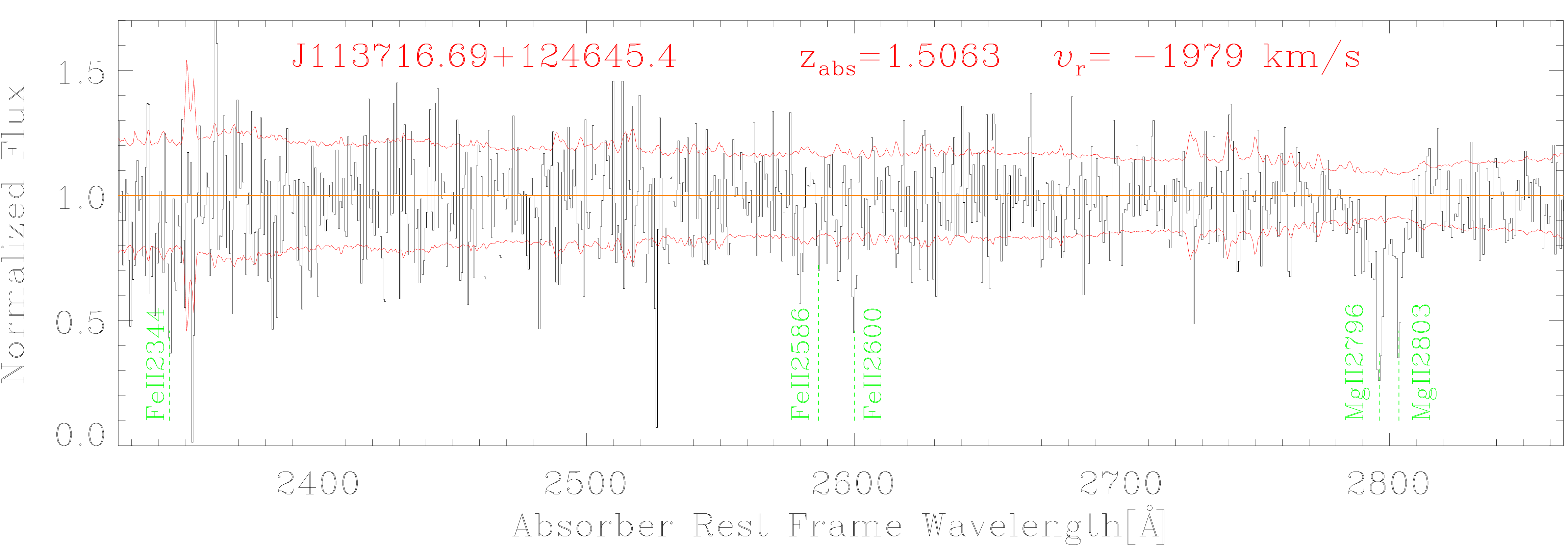}\\
\includegraphics[width=12cm,height=2.6cm]{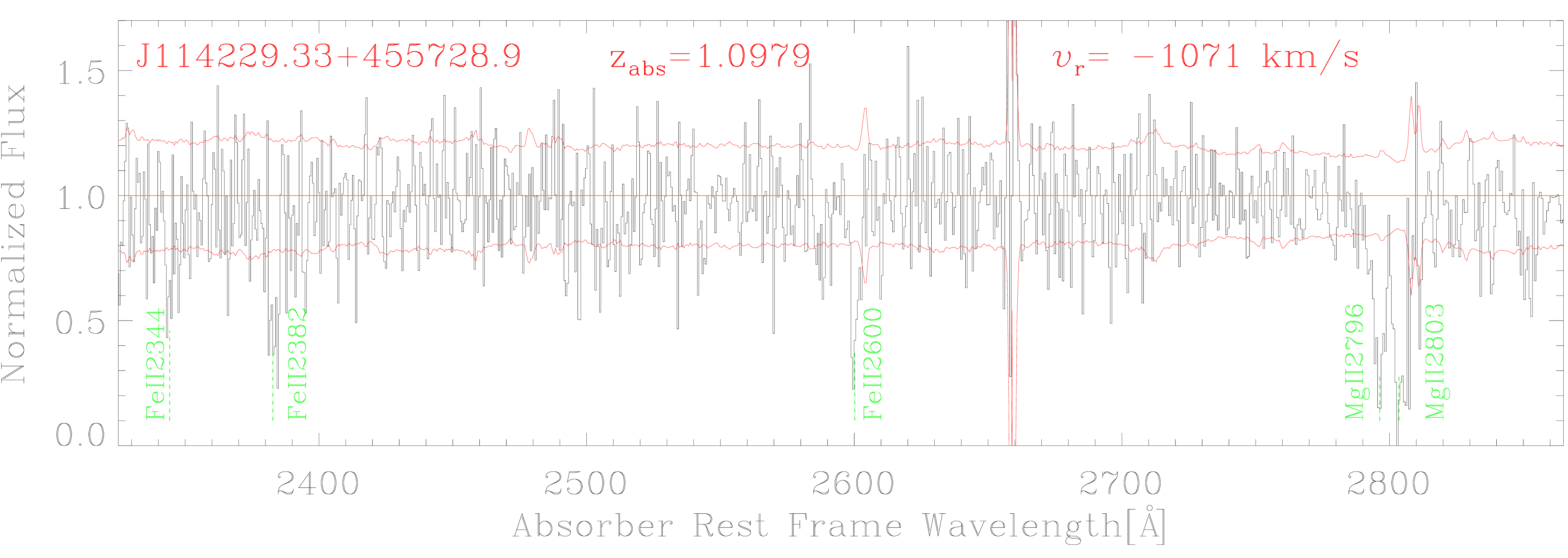}\\
\includegraphics[width=12cm,height=2.6cm]{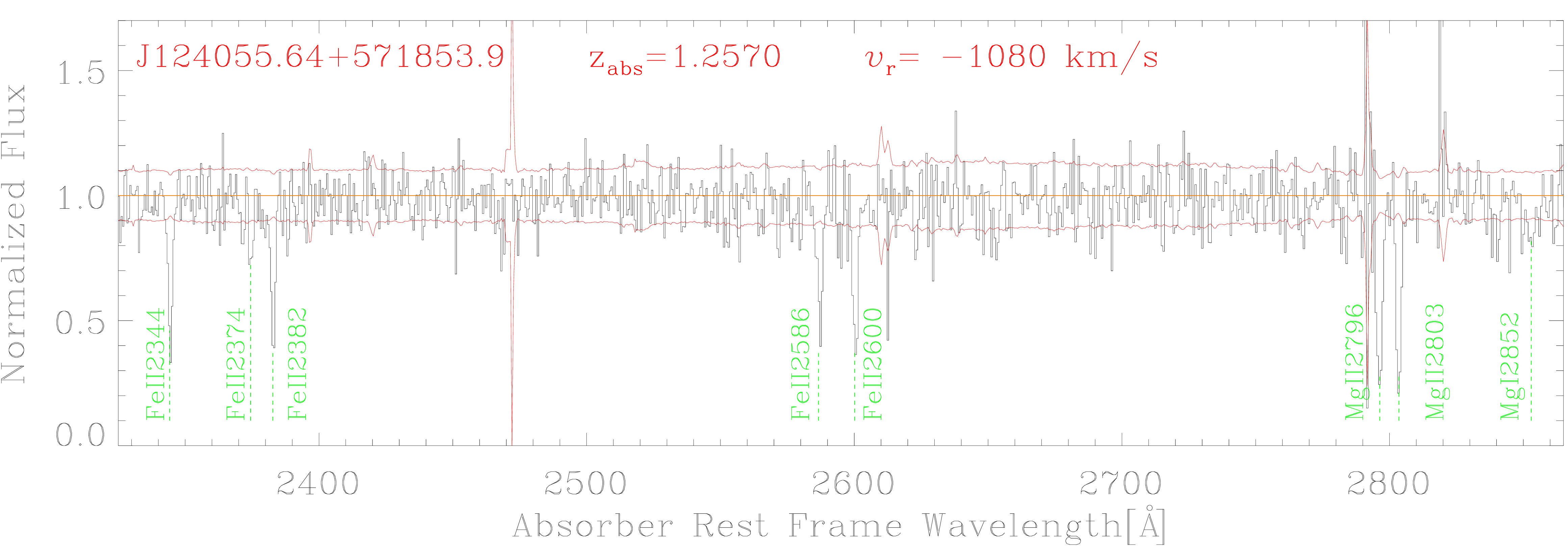}\\
\includegraphics[width=12cm,height=2.6cm]{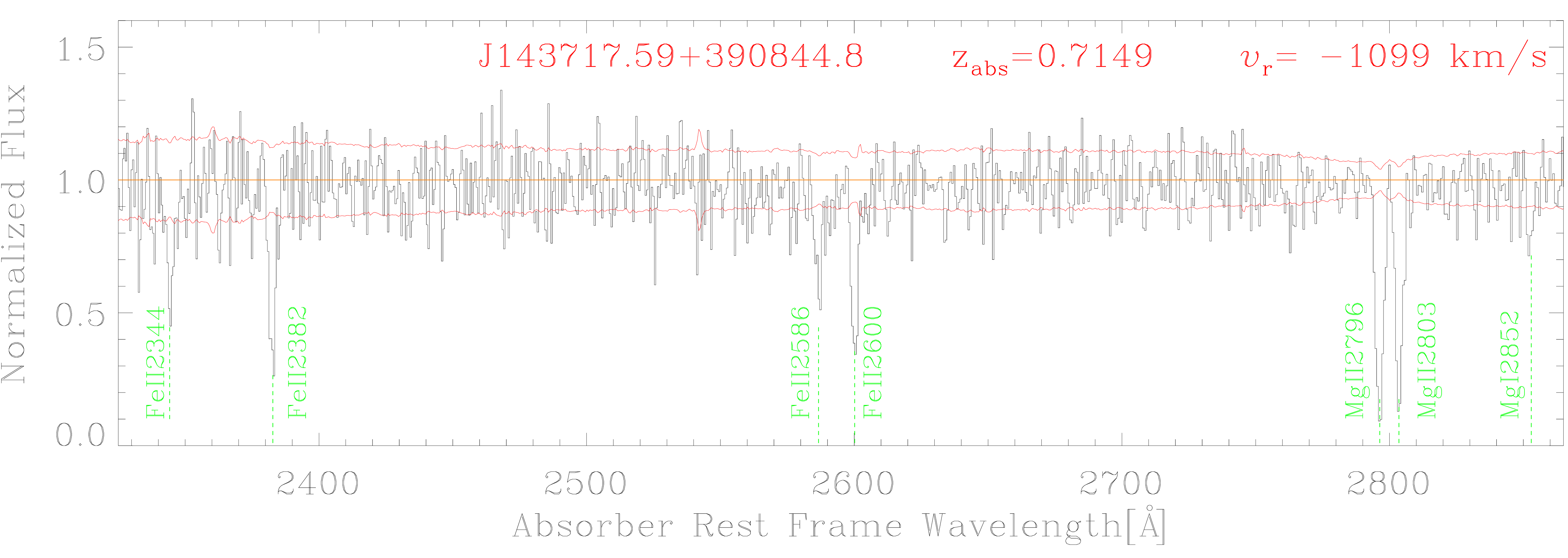}\\
\includegraphics[width=12cm,height=2.6cm]{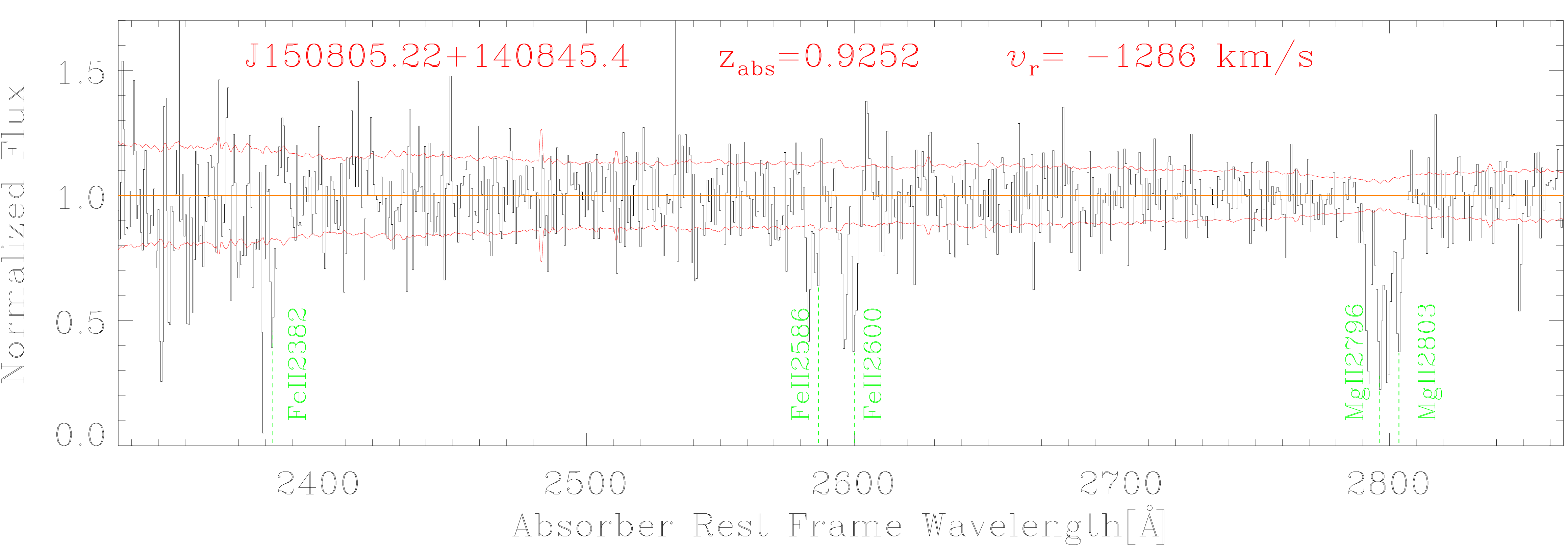}\\
\includegraphics[width=12cm,height=2.6cm]{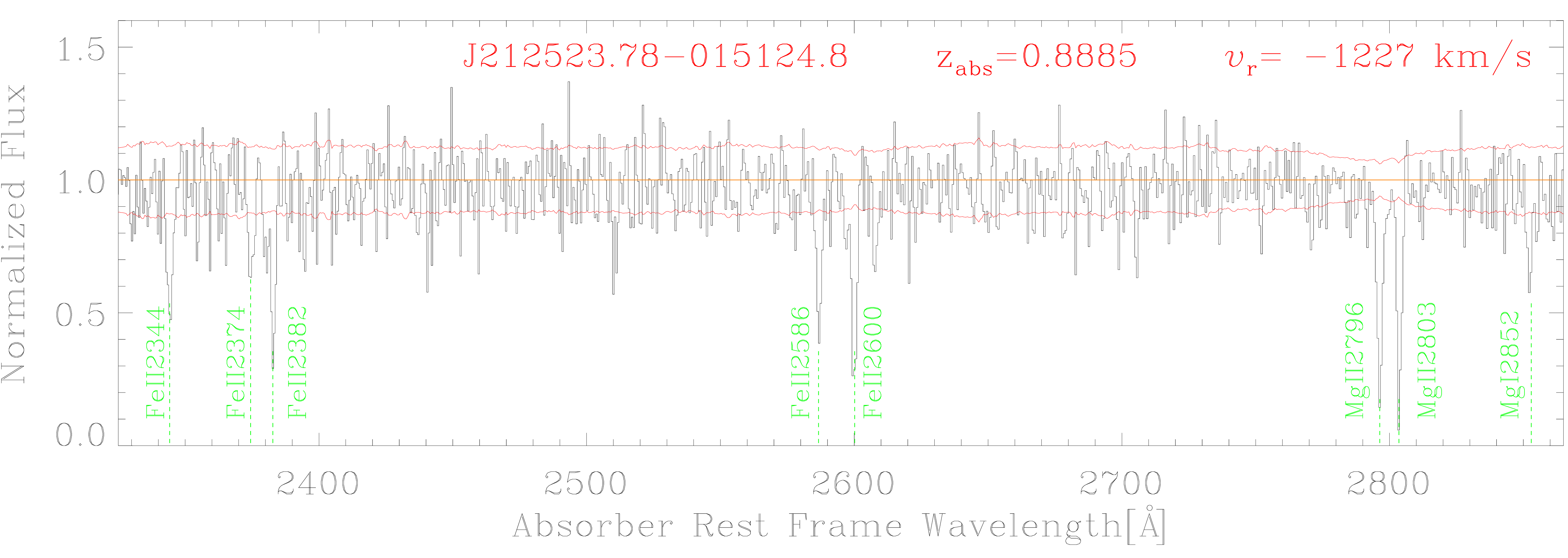}\\
\caption{The quasar spectra normalized by fitting continuum. The red lines are the $\pm1\sigma$ flux uncertainty levels, which have also been normalized by the fitting continuum. Green dash-lines label the obvious absorption lines located at the same $z_{\rm abs}$, where the values of $z_{\rm abs}$ are in the top of each figure.}
\label{fig:abs}
\end{figure*}

In term of the discussions mentioned above, the 9 redshifted \MgII\ NALs with $\upsilon_r<-1000$ \kms\ are likely originated within the inflowing material. We show the spectra of the 9 redshifted \MgII\ NALs with $\upsilon_r<-1000$ \kms\ in Figures \ref{fig:qso} and \ref{fig:abs}. For all these 9 redshifted \MgII\ NALs with $\upsilon_r<-1000$ \kms, we also measure the absorption strengths of the \MgI, the series of \FeII, and the \CIVab\ when available. The results are listed in Table \ref{Tab:abs}.

\begin{table*}
\caption{The measurements of absorption lines} \tabcolsep 1mm \centering \scriptsize
\label{Tab:abs}
 \begin{tabular}{cccccccccccccccccc}
 \hline\hline\noalign{\smallskip}
SDSS NAME  & \zabs & $W_{\rm r}^{\lambda2796}$ & $W_{\rm r}^{\lambda2803}$ &  $W_{\rm r}^{\lambda2852}$ & $W_{\rm r}^{\lambda2600}$ & $W_{\rm r}^{\lambda2586}$ & $W_{\rm r}^{\lambda2382}$ & $W_{\rm r}^{\lambda2374}$ & $W_{\rm r}^{\lambda2344}$ & $W_{\rm r}^{\lambda1548}$ & $W_{\rm r}^{\lambda1551}$  \\
&&\AA&\AA&\AA&\AA&\AA&\AA&\AA&\AA&\AA&\AA \\
(1)&(2)&(3)&(4)&(5)&(6)&(7)&(8)&(9)&(10)&(11)&(12)\\
\hline\noalign{\smallskip}
004039.79+150321.2 & 0.8929 & 0.65$\pm$0.07 & 0.69$\pm$0.08 & 0.42$\pm$0.18 & 0.36$\pm$0.09 & 0.68$\pm$0.20 & 0.44$\pm$0.17 &---&---&---&---  \\
021144.19-063726.8 & 1.0278 & 1.13$\pm$0.13 & 0.98$\pm$0.16 & 0.32$\pm$0.10 & 0.45$\pm$0.13 & 0.41$\pm$0.10 & 0.72$\pm$0.18 &---& 0.38$\pm$0.09 & ---&--- \\
111537.58+121845.2 & 1.1868 & 1.47$\pm$0.07 & 1.17$\pm$0.07 & 0.68$\pm$0.09 & 1.15$\pm$0.10 & 0.68$\pm$0.09 & 1.00$\pm$0.08 & 0.40$\pm$0.09 & 0.93$\pm$0.09 &---&---  \\
113716.69+124645.4 & 1.5063 & 2.02$\pm$0.22 & 1.49$\pm$0.21 &---            & 1.06$\pm$0.31 & 0.41$\pm$0.20 &---            &---& 1.01$\pm$0.33 & 0.57$\pm$0.17 & 0.22$\pm$0.10 \\
114229.33+455728.9 & 1.0979 & 3.99$\pm$0.69 & 4.64$\pm$0.75 &---            & 2.51$\pm$0.72 &---            & 2.90$\pm$0.94 &---& 1.30$\pm$0.51 &---&---\\
124055.64+571853.9 & 1.2570 & 2.04$\pm$0.20 & 1.86$\pm$0.19 & 0.57$\pm$0.22 & 1.58$\pm$0.28 & 1.28$\pm$0.27 & 1.29$\pm$0.19 & 0.57$\pm$0.19 & 1.27$\pm$0.18 &---&---  \\
143717.59+390844.8 & 0.7149 & 3.40$\pm$0.21 & 2.88$\pm$0.21 & 0.72$\pm$0.26 & 2.20$\pm$0.32 & 1.34$\pm$0.29 & 2.25$\pm$0.42 &---& 1.50$\pm$0.43 & ---&--- \\
150805.22+140845.4 & 0.9252 & 1.92$\pm$0.14 & 1.54$\pm$0.17 &---            & 1.42$\pm$0.27 & 0.76$\pm$0.30 & 1.07$\pm$0.31 &---&---&---&---\\
212523.78-015124.8 & 0.8885 & 2.24$\pm$0.19 & 1.99$\pm$0.17 & 1.09$\pm$0.36 & 2.02$\pm$0.27 & 1.48$\pm$0.29 & 1.79$\pm$0.35 & 0.98$\pm$0.30 & 1.38$\pm$0.31 &---&--- \\
\hline\hline\noalign{\smallskip}
\end{tabular}
\begin{flushleft}
\textbf{Note.} (1) The SDSS name of quasars; (2) The redshift of \MgII\ NALs; (3)---(12) The equivalent widths of \MgIIab, \MgI, \fea, \feb, \fec, \fed, \fee, \CIVab\ absorption lines. The ``---'' means no detection.
\end{flushleft}
\end{table*}

In order to measure the black hole mass and bolometric luminosities of quasars, we further analyze the quasar spectra. All the quasar spectra are corrected for the Galactic extinction using the Reddening measurements of \citet{2011ApJ...737..103S} and Milky Way extinction curve from \citet{1989ApJ...345..245C}. We fit local power-law continuum ($f_{\lambda}=A\lambda^{\alpha}$) plus iron template \citep[][]{2001ApJS..134....1V} using the spectra data around the \MgII\ emission lines and not contaminated by the considering line emissions. See Figures \ref{fig:qso} for the fitting results. The \MgII\ emission line properties are measured from the spectra subtracted by the continuum+iron fits. We invoked two Gaussian functions to fit the \MgII\ emission features, of which one having FWHM $>1200$ \kms\ \citep[e.g.,][]{2011ApJS..194...45S,2019ApJS..244...36C} is for the broad \MgII\ component, and the other one, whose FWHM is determined from the \OII\ emission lines, is for the narrow \MgII\ component. During the fits of emission lines, we exclude the spectral data around the strong \MgIIab\ absorption doublets. We derive the black hole mass $M_{\rm BH}$ of the quasar via
\begin{equation}\label{eq:MBH}
 \rm Log (\frac{M_{BH}}{M_\odot}) = a + b\times Log(\frac{L_{3000}}{10^{44}~erg~s^{-1}}) + 2\times Log(\frac{FWHM_{Mg~II}}{km~s^{-1}}),
\end{equation}
where the calibrated coefficients $(a,b)=(0.82,0.5)$ are empirical values \cite[e.g.,][]{2019ApJS..244...36C}, and the $L_{\rm 3000}$ is the monochromatic luminosity of the quasar at 3000 \AA, which is estimated from the fitting power-law. The bolometric luminosity of quasar is directly derived from the monochromatic luminosity with standard bolometric correction, namely, $L_{\rm bol} = 5.15 L_{\rm 3000}$ \citep[][]{2006ApJS..166..470R}. We also estimate the mass accretion rate of black hole by $\dot{M}_{\rm BH} = \frac{L_{\rm bol}}{\epsilon c^2}$, where $\epsilon = 0.1$ is the mass-to-radiation conversion efficiency. The Eddington ratio can be derived from $\eta = L_{\rm bol}/L_{\rm Edd}$, where the Eddington luminosity $L_{\rm Edd} = 1.38\times10^{38}(M_{\rm BH}/M_{\bigodot})$ \ergs. The resulting $L_{\rm 3000}$, $M_{\rm BH}$, $\dot{M}_{\rm BH}$, and $\eta$ are provided in Table \ref{Tab:qso}.

\begin{table*}
\caption{The parameters of quasars and absorption line systems} \tabcolsep 1.mm \centering \footnotesize
\label{Tab:qso}
 \begin{tabular}{cccccccccccccccccccccc}
 \hline\hline\noalign{\smallskip}
SDSS NAME         & \zem    & \zabs   & $\upsilon_{r}$ & R &$C_{\rm f}$&$Log L_{\rm 3000}$ & $Log M_{\rm BH}$ & $\eta$ & $\rm \dot{M}_{\rm BH}$ & $R_{\rm BELR}$ & $R_{\rm torus}$  & $R_{\rm NELR}$ & $R_{\rm abs}^{\rm HG}$ & $R_{\rm abs}^{\rm BH}$ \\
{\smallskip}
&           &          &    \kms   & &&  \ergs   & $\rm M_{\bigodot}$&   &  $\rm M_{\bigodot}/yr$ &     pc    & pc& pc& pc& pc       \\
{\smallskip}
(1) &        (2)      &    (3)    &    (4)  &  (5)  &    (6)         &   (7)  &    (8) &   (9) &   (10) &  (11) &  (12)  &  (13) & (14)   & (15)       \\
\hline\noalign{\smallskip}
004039.79+150321.2 &  0.8857   &  0.8929   &  -1148 &  753   &  0.52   &  45.33   &  9.06   &  -1.131   &  1.96   &  0.126   &  1.595   &  493.30  &  12.46 & 7.53   	\\
021144.19-063726.8 &  1.0205   &  1.0278   &  -1081 &  ---   &  0.53   &  45.16   &  8.84   &  -1.081   &  1.31   &  0.144   &  1.304   &  376.59  &  7.26  & 5.06   \\
111537.58+121845.2 &  1.1790   &  1.1868   &  -1073 &  ---   &  0.60   &  45.79   &  9.08   &  -0.698   &  5.53   &  0.219   &  2.682   &  990.42  &  16.05 & 8.99   \\
113716.69+124645.4 &  1.4898   &  1.5063   &  -1979 &  5703  &  0.81   &  45.19   &  8.95   &  -1.170   &  1.37   &  0.104   &  1.333   &  388.21  &  1.48  & 1.94   \\
114229.33+455728.9 &  1.0904   &  1.0979   &  -1071 &  546   &  0.84   &  44.85   &  8.81   &  -1.367   &  0.64   &  0.069   &  0.909   &  232.25  &  6.67  & 4.83   \\
124055.64+571853.9 &  1.2489   &  1.2570   &  -1080 &  79    &  0.82   &  45.43   &  8.89   &  -0.862   &  2.46   &  0.142   &  1.788   &  575.15  &  8.04  & 5.76   \\
143717.59+390844.8 &  0.7086   &  0.7149   &  -1099 &  ---   &  0.93   &  44.60   &  8.33   &  -1.129   &  0.36   &  0.054   &  0.684   &  158.76  &  1.70  & 1.50   \\
150805.22+140845.4 &  0.9170   &  0.9252   &  -1286 &  7717  &  0.77   &  45.42   &  8.70   &  -0.689   &  2.37   &  0.140   &  1.755   &  560.78  &  2.78  & 2.62   \\
212523.78-015124.8 &  0.8808   &  0.8885   &  -1227 &  ---   &  0.83   &  44.57   &  8.28   &  -1.123   &  0.34   &  0.042   &  0.662   &  151.97  &  1.04  & 1.09   \\
\hline\hline\noalign{\smallskip}
\end{tabular}
\begin{flushleft}
\textbf{Note.} Column (4): Relative velocities of absorbers with respect to quasar emission line redshifts; (5) $R = \frac{f_{6cm}}{f_{2500}}$, which represents the radio loudness of quasars; (6) The coverage fraction of the \MgII\ absorbing cloud relative to the continuum emission region; (7) The monochromatic luminosity of the quasar at 3000 \AA; (8) The mass of the black hole; (9) The Eddingtion ratio $\eta = L_{\rm bol}/L_{\rm Edd}$; (10) The mass accretion rate of the black hole; (11) The size of the broad emission line region; (12) The inner radius of dusty torus; (13) The size of the narrow emission line region; (14) The largest distances of absorbers from central SMBH, considering the gravitational potential contributed by the central supermassive black hole and a Hernquist stellar distribution for the hosting galaxy \citep{1990ApJ...356..359H}; (15) The largest distances of absorbers from central SMBH, only considering the gravitational field of central SMBH.
\end{flushleft}
\end{table*}

\section{The results and discussions}
In the following, we discuss the properties of locations, mass inflow rates, radio emissions, and origins for the 9 redshifted \MgII\ NALs with $\upsilon_r<-1000$ \kms.

\subsection{The locations of infalling absorbers}
We estimate the locations ($R_{\rm abs}^{\rm HG}$) of these inflowing NALs assuming that they are initially formed in the hosting galaxies and then fall freely toward the center. Due to the drag force from the surrounding hot gas, the centrifugal force and the potential radiation pressure from the central source, the true line-of-sight velocities of infalling NALs should be less than the free-fall velocities. Therefore, our estimated values are an upper limit for $R_{\rm abs}^{\rm HG}$. We consider the gravitational potential contributed by the central supermassive black hole and a Hernquist stellar distribution for the bulge of the hosting galaxy \citep{1990ApJ...356..359H}. The bulge stellar mass $M_{*}$ is derived from the black hole mass $M_{\rm BH}$ from the $M_ {\rm BH}-M_{*}$ relation given by \citet{2004ApJ...604L..89H},
\begin{eqnarray}
\frac{M_{\rm BH}}{M_{\sun}  }= 1.6\times10^{8}\left(\frac{M_{*}}{10^{11}M_{\sun}}\right)^{1.12}   \text{,}
\end{eqnarray}
\noindent
and the effective radius $R_{e}$ of the bulge stellar distribution is then determined from the observational $R_{e}-M_ {*}$ relation for early-type galaxies in \citet{2003MNRAS.343..978S},
\begin{eqnarray}
R_{e}= 2.88\times 10^{-6}\left(\frac{M_{*}}{M_{\sun}}\right)^{0.56}     \text{~kpc.}
\end{eqnarray}
We then calculate the values of $R_{\rm abs}^{\rm HG}$ assuming that the free-fall velocities of the gas clouds reach the observed inflowing velocities at $R_{\rm abs}^{\rm HG}$ in this potential well. The resulting distances of our inflowing absorbers $R_{\rm abs}^{\rm HG}$ are provided in Table \ref{Tab:qso}. We find that the derived values of $R_{\rm abs}^{\rm HG}$ for our 9 infalling \MgII\ NALs are all roughly $2\times 10^4 R_g$ --- $2\times10^5 R_g$, substantially lower than the radius of the sphere of the influence of the central SMBHs \citep[$\sim~10^6 R_{\rm g}$ assuming a representative bulge stellar velocity dispersion of $\sim 200$ km/s,][]{2015ARA&A..53..365N}.

If considering the inflowing NALs are initially formed around dusty tori, the infalling gas clouds are dominated by the gravitational field of the central SMBH. In this case, the resulting distances of inflowing absorbers $R_{\rm abs}^{\rm BH}$ are lightly smaller than the $R_{\rm abs}^{\rm HG}$. The resulting $R_{\rm abs}^{\rm BH}$ are also listed in Table \ref{Tab:qso}.

\begin{figure}
\centering
\includegraphics[width=0.43\textwidth]{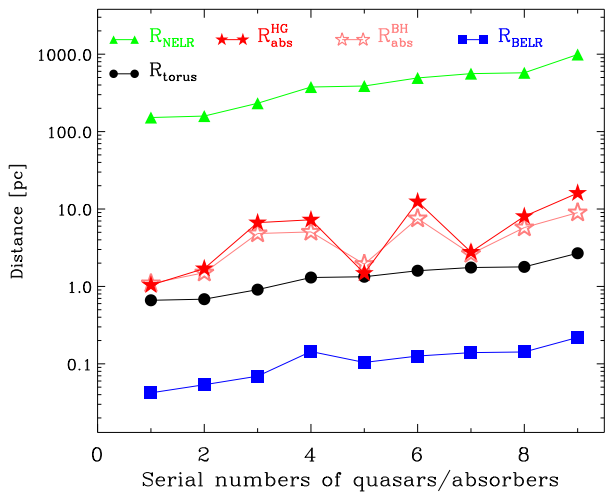}\\
\caption{The distances of the broad emission line region ($R_{\rm BELR}$), dusty torus ($R_{\rm torus}$), narrow emission line region ($R_{\rm NELR}$), and inflow cloud ($R_{\rm abs}^{\rm HG}$ or $R_{\rm abs}^{\rm BH}$) from the quasar central region. The $R_{\rm abs}^{\rm HG}$ considers the gravitational potential contributed by the central supermassive black hole and a Hernquist stellar distribution for the bulge of the hosting galaxy \citep{1990ApJ...356..359H}, and the $R_{\rm abs}^{\rm BH}$ only considers the gravitational field of central SMBH.}
\label{fig:dist}
\end{figure}

Using the radius-luminosity relations, we derive the sizes of broad emission line regions \citep[$R_{\rm BELR}$,][]{2013ApJ...767..149B} and narrow emission line regions \citep[$R_{\rm NELR}$,][]{2012MNRAS.420..526M}, and the inner radius of dusty torus \citep[$R_{\rm torus}$,][]{2008ApJ...685..160N} from quasar luminosities. The resulting in $R_{\rm BELR}$, $R_{\rm NELR}$, and $R_{\rm torus}$ are shown in Figure \ref{fig:dist} and also listed in Table \ref{Tab:qso}. It is clearly seen from Figure \ref{fig:dist} that the $R_{\rm abs}$ is just slightly larger than the inner radius of dusty torus $R_{\rm torus}$, implying that the infalling \MgII\ NALs could be formed within the gas clouds located near the dusty tori.

\subsection{The mass inflow rates}
Assuming a simplest model that the infalling \MgII\ NALs is formed within a discrete clouds \cite[][]{2019Natur.573...83Z}, we can estimate the mass inflow rate via
\begin{equation}\label{eq:Minflow}
\dot{M}_{inflow} = \mu m_{\rm p} N_{\rm H,inflow} f_{\rm f} 4\pi d_{\rm inflow}^2 \Omega_{\rm inflow} / t_{\rm inflow}
\end{equation}
where $\mu = 1.4$ is the mean atomic mass per proton, $m_{\rm p}$ is the mass of a proton, $N_{\rm H,inflow}$ is the column density of neutral hydrogen within the inflow cloud, $f_{\rm f}$ is the filling factor of the absorbing clouds, the in-falling distance $d_{\rm inflow}$ equals to the distance $R_{\rm abs}$ of the absorbing cloud, $\Omega_{\rm inflow}$ is the global covering factor of the inflow structure, and $t_{\rm inflow}$ is the in-falling timescale. Considering the gravitation and radiation pressure from central region, the infalling timescale of the inflow absorber $t_{\rm inflow} \approx d_{\rm inflow}/\upsilon_{\rm inflow}$ when the absorber is falling towards central black hole from the observed distance, where $\upsilon_{\rm inflow}$ equals to the relative velocity $\upsilon_r$ of the \MgII\ NALs measured from quasar spectra.
The filling factor $f_{\rm f}$ may be the same order as the effective coverage fraction $C_{\rm f}$ of the absorbing cloud to background emission sources for the BALs \cite[][]{2019Natur.573...83Z}. However, the $f_{\rm f}$ may be much smaller than the $C_{\rm f}$ for the NALs. In this paper, we only estimate the upper limit of the mass inflow rate from Equation (\ref{eq:Minflow}), therefore, we still consider $f_{\rm f}=C_{\rm f}$. This would not change our result.

Among 9 redshifted \MgII\ NALs with $\upsilon_r<1000$ \kms, the high ionization \CIV\ absorption is only available for one system, and we only detect the $\rm Mg^+$ and $\rm Fe^+$ iron absorption lines for the other 8 systems with available spectra. The hydrogen column density $(N_{\rm H})$ of the inflowing absorbers cannot be well determined by the photoionization simulations \citep[][]{2017RMxAA..53..385F} if the $\rm Mg^+$ and $\rm Fe^+$ iron absorption lines are only available, since both the $\rm Mg^+$ and $\rm Fe^+$ irons have similar ionization potential. As a matter of fact, the metal BALs and DLAs (damped $Ly\alpha$ absorption line systems) often host a hydrogen column density significantly larger than $10^{20}~cm^{-2}$, and the metal NALs usually have $N_{\rm H}<10^{20}~cm^{-2}$. Here we assume $N_{\rm H}=10^{20}~cm^{-2}$ for all the 9 inflowing \MgII\ NALs to estimate the upper limit of mass inflow rates.

\begin{figure}
\centering
\includegraphics[width=0.43\textwidth]{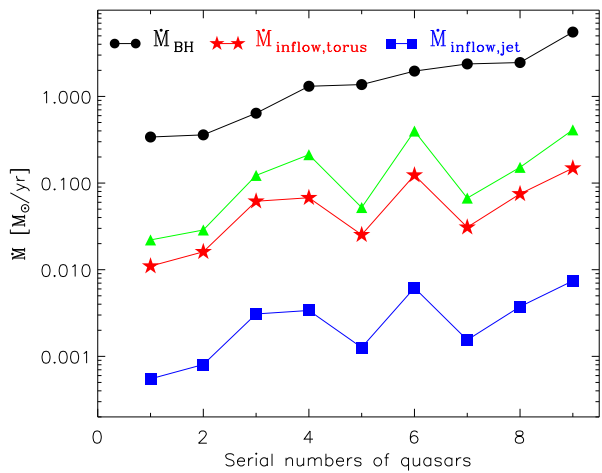}\\
\caption{The mass accretion rate of the central black hole (black circles) and the mass inflow rate of absorbing gas (color symbols). Red stars are for $\Omega_{\rm inflow}=0.6$ (dusty torus), blue squares are for $\Omega_{\rm inflow}=0.03$ (back flow), and green triangles are for $\Omega_{\rm inflow}=1$ and $C_f=1$.}
\label{fig:Mar}
\end{figure}

The absorption line intensity ($I(\upsilon)$), which has been normalized by the background radiations, is a function of the optical depth ($\tau(\upsilon)$) and the effective coverage fraction ($C_{\rm f}(\upsilon)$) of the absorbing cloud to background emission sources. That is,
\begin{equation}\label{eq:Iv}
I(\upsilon)=[1-C_{\rm f}(\upsilon)]+C_{\rm f}(\upsilon)e^{\rm -\tau(\upsilon)}.
\end{equation}
For the \MgIIab\ doublet that has an optical depth ratio of $2:1$, we assume that $C_{\rm f}$ is the same for both lines. Therefore, we can obtain
\begin{equation}\label{eq:cf}
C_{\rm f}(\upsilon)=\frac{[I_{\rm r}(\upsilon)-1]^2}{I_{\rm b}(\upsilon)-2I_{\rm r}(\upsilon)+1}.
\end{equation}
Considering the resolution of the SDSS spectra $R=\Delta \lambda/\lambda\approx1800$ and line profile of narrow absorption line, we only derive the effective coverage fraction of the absorbing cloud using the normalized intensities at line cores. The results are provided in Table \ref{Tab:qso}. Note that the $C_{\rm f}$ is only used to estimate the mass inflow rate (see Equation (\ref{eq:Minflow})). In the following comparison (see Figure \ref{fig:Mar}), we will see that an accurate $C_{\rm f}$ cannot change the results, and the $\dot{M}_{inflow}$ is still much less than the $\dot{M}_{\rm BH}$, even if we adopt $C_{\rm f}=1$.

With the $N_{\rm H}=10^{20}~cm^{-2}$, $C_{\rm f}$, $R_{\rm abs}$, and $\upsilon_r$ being available/given, we can obtain $\rm \dot{M}_{\rm inflow}$=(0.0183 --- 0.2478)$\Omega_{\rm inflow}~M_{\bigodot}/yr$ for our 9 infalling \MgII\ NALs from Equation \ref{eq:Minflow}, where the $\Omega_{\rm inflow}<1$. Figure \ref{fig:Mar} compares the mass accretion rate of black hole and the mass inflow rate of absorbing gas for different $\Omega_{\rm inflow}$ values. It is clearly seen that $\dot{M}_{inflow}\ll\dot{M}_{\rm BH}$, suggesting that the infalling NALs cannot provide sufficient fuels to power the quasars.

\subsection{The properties of Radio emissions}\label{Sect:radio}
We collect quasar emission data at radio-bands through the High Energy Astrophysics Science Archive Research Center Online Service, provided by the NASA/Goddard Space Flight Center. We find that all of these 9 quasars are lied within the footprint of the Faint Images of the Radio Sky at Twenty cm (FIRST). Therefore, firstly, we match our quasar sample to FIRST using a $5''$ matching radius, which results in 5 quasars with the FIRST detections. Secondly, we match the 5 quasars with the FIRST detections to other catalogs of radio observations using a $60''$ matching radius. We find that, except for the 20 cm observations, there are 4 quasars being observed at other radio frequencies. For the quasars with the FIRST detection within a $5''$ matching radius, we also match the quasars with the FIRST catalog using a $60''$ matching radius. Within a $60''$ matching radius, the quasars only with one counterpart of the FIRST are likely to be core-dominated sources, and those with at least two counterparts of the FIRST are likely to be lobe-dominated sources \cite[e.g.,][]{2011ApJS..194...45S}. Based on this criterion, we find that 3 quasars are core-dominated sources among the 5 quasars with the FIRST detections. This indicates that most of the quasars with the FIRST detections are likely observed along the directions close to the axes of accretion disks (with small inclination angles), which are the directions at which the NALs associated with quasar outflows/winds were observed \cite[][]{2012ASPC..460...47H}.

We also compute the radio loudness of quasars via
\begin{equation}\label{eq:loudness}
R = \frac{f_{6~cm}}{f_{2500}},
\end{equation}
where the $f_{6~cm}$ and $f_{2500}$ are the flux density ($f_{\nu}$) at 6 cm and 2500 \AA, respectively. The $f_{2500}$ is directly taken from the fitting pow-law continuum of quasars at 2500 \AA. The flux density at rest-frame 6 cm is estimated from multi-band measurements using the fitted power-law spectral index, otherwise is from the only available FIRST 20 cm observations assuming a spectral index of 0.5. During the power-law fits in radio-bands, we adopt the total flux of all the components for the two lobe-dominated sources. We note that the observations at different radio bands were done at different resolutions, which may yield a different flux at the same frequency for the same one source, especially for the extended sources. Except for the FIRST detection (at a resolution of $5''$), the radio observations from other surveys were done at much lower resolutions ($\ge45''$). Here, we compare the fluxes at 1.4 GHz obtained by the FIRST (at a resolution of $5''$) and the National Radio Astronomy Observatory Very Large Array Sky Survey \cite[NVSS,][at a resolution of $45''$]{1998AJ....115.1693C}, respectively. We find that the fluxes obtained by the FIRST are similar to those obtained by the NVSS, which indicates that the resolutions of observations would not yield significant difference in the calculations of radio indices.

The resulting in $R$ from Equation \ref{eq:loudness} are listed in Table \ref{Tab:qso}. We find that $R=79$, 546, 753, 5703, and 7717 for the 5 quasars with the FIRST detections, which are within the range covered by blazars \citep[e.g.,][]{2011RAA....11.1413F,2019RAA....19...70P,2020RAA....20...25P}.
The radio loudness of quasars suggests that the 5 quasars with the FIRST detected are likely Blazars, which further supports that these sources are likely observed along the directions close to the axes of accretion disks.

\subsection{The Eddington ratios}\label{Sect:radio}
\begin{figure}
\centering
\includegraphics[width=0.48\textwidth]{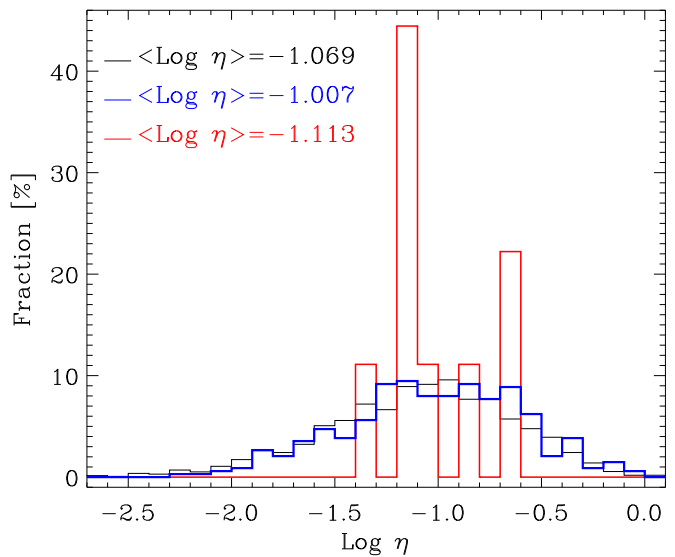}\\
\caption{Distributions of the Eddington ratios $\eta$ of quasars. The black solid-line is for all the quasars with $z_{\rm em}<1.07$ (2724 quasars), whose black hole mass can be available from previous works. The median value is $\rm \langle Log \eta \rangle=-1.069$. The blue solid-line is for the quasars with outflow \MgII\ NALs (338 quasars), whose median value is $\rm \langle Log \eta \rangle=-1.007$. The red solid-line is for the 9 quasars with inflowing \MgII\ NALs, whose median value is $\rm \langle Log \eta \rangle=-1.113$.}
\label{fig:Eddington}
\end{figure}

The Eddington ratios of the 9 quasars with inflowing \MgII\ NALs are shown with red solid-line in Figure \ref{fig:Eddington}. We calculate the Eddingtong ratios for all the quasars with $z_{\rm em}<1.07$, included in \citet{2018ApJS..235...11C}, when their black hole mass can be available from previous works \cite[e.g.,][]{2011ApJS..194...45S,2018ApJS..234...16C,2019ApJS..244...36C}. The results are shown with black solid-line in Figure \ref{fig:Eddington}, which includes 2724 quasars. It is clearly seen from Figure \ref{fig:dopp} that the \MgII\ NALs with $3*279 < \upsilon_r < 1549 + 848$ \kms\ would be dominated by outflow absorptions, where the 279 \kms\ is the velocity dispersion of the environmental absorptions (the blue dash-line in Figure \ref{fig:dopp}), the 1548 \kms\ is the central velocity of the outflow absorptions (the blue solid-line in Figure \ref{fig:dopp}), and the 848 \kms\ is the velocity dispersion of the outflow absorptions. There are 338 quasars, whose \MgII\ NALs fall into the velocity range of $3*279 < \upsilon_r < 1549 + 848$ \kms. Here these 338 quasars are considered as the sources with outflow \MgII\ NALs, whose Eddington ratios are shown with blue solid-line in Figure \ref{fig:Eddington}. It is clearly seen from \ref{fig:Eddington} that, on average, the quasars with inflowing \MgII\ NALs have a slightly smaller Eddington ratio relative to those with outflow \MgII\ NALs. It can be seen from Figure \ref{fig:dopp} that the outflow component (blue solid-line) is significantly contaminated by the intervening absorptions (green solid-line) in the velocity range of $3*279 < \upsilon_r < 1549 + 848$ \kms. Therefore, above definition of quasars with outflow \MgII\ NALs would significantly reduce the purity of sample of quasars with outflow \MgII\ NALs. The Eddington ratios of the quasars with confirmed outflow \MgII\ NALs might be on average larger than those shown with blue solid-line in Figure \ref{fig:Eddington}. Thus, Figure \ref{fig:Eddington} provides a hint that the quasars with inflowing \MgII\ NALs may have a lower radiation pressure with respect to the sources with outflow \MgII\ NALs. In the future, we will further investigate whether the properties of inflow/outflow \MgII\ NALs depend on the intrinsic characteristics of quasars, when a large sample of quasars with confirmed outflow \MgII\ NALs is available.

\section{The origins of infalling \MgII\ NALs}
\begin{figure*}
\centering
\includegraphics[width=0.45\textwidth]{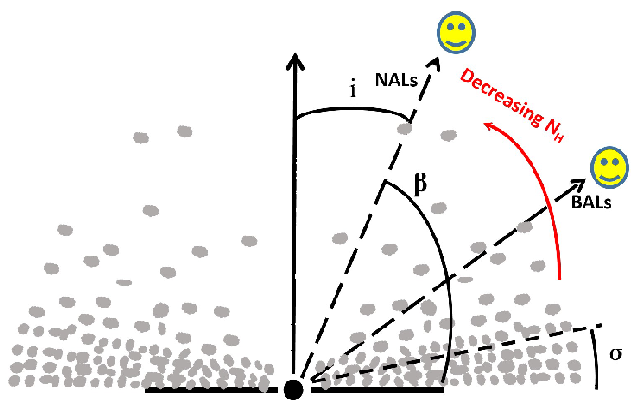}
\hspace{5ex}
\includegraphics[width=7.5cm,height=5.5cm]{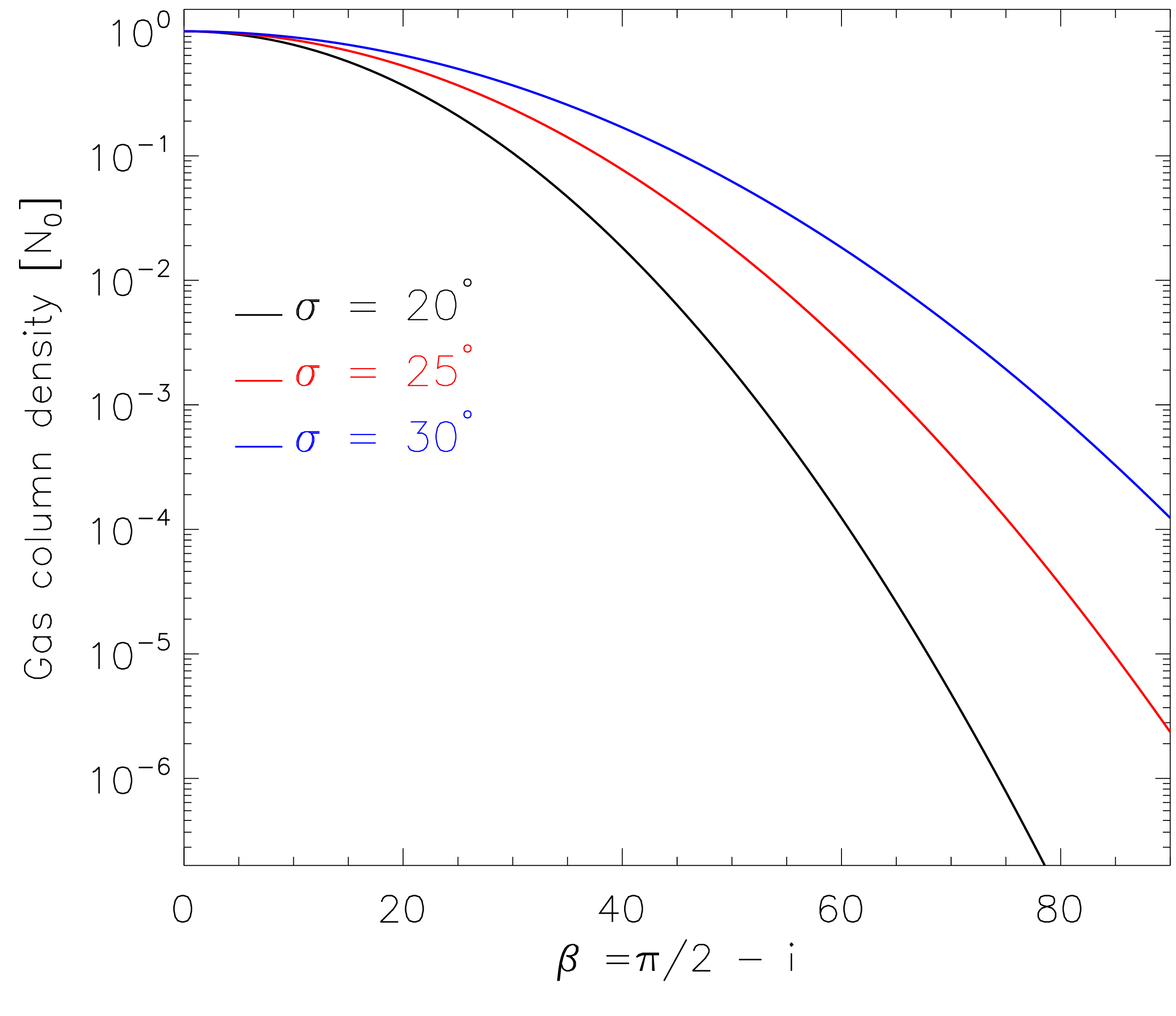}\\
\caption{Left panel: A cartoon for the distribution of the absorbing gas from accretion disk plane. $i$ is the angle between sightline and axial direction of accretion disk, and $\beta = \frac{1}{2}\pi - i$. The gas column density is a constant within the angle width $\mid\beta\mid\leq\sigma$. The BALs would be possibly observed along sightline with small $\beta$, and the NALs would be possibly detected along the sightline with small $i$. Right panel: The gas column density as a function of $\beta$, giving an angle width $\sigma$.}
\label{fig:carton}
\end{figure*}

The upper limits of $R_{\rm abs}$ are around the regions of dusty tori. Therefore, one possible origin of the infalling \MgII\ NALs is that they are from dusty tori. Within the dusty tori, clouds have a Gaussian distribution along the angular direction, and is decreased in the way of $1/r$ or $1/r^2$ along the radial direction \citep[][]{2008ApJ...685..160N}. Therefore, we assume a Gaussian distribution of the absorbing gas from accretion disk plane, when the infalling \MgII\ NALs are formed within the dusty tori. That is,
\begin{equation}\label{eq:cartoon}
N_{\rm g}(\beta) = N_{\rm 0} e^{(-\beta^2/\sigma^2)},
\end{equation}
where $N_{\rm g}$ is the column density of gas cloud, $N_{\rm 0}$ is a normalized parameter, $\beta$ is the angle between sightline and accretion disk plane, $\sigma$ is a typical angle width of gas distribution, and the gas column density is a constant within the angle width $\mid\beta\mid\leq\sigma$, namely, $N_{\rm g} = N_{\rm 0}$. Figure \ref{fig:carton} shows the distribution of gas column density as a function of $\beta$ for several $\sigma$. In this scenario, the BALs and NALs are observed along sightlines with small and large $\beta$, respectively, which are consistent with the model of \citet{2012ASPC..460...47H}. Assuming that $\sigma =20^\circ$ and the infalling BALs of quasar J103516.20 + 142200.6 \citep[$\rm N_{\rm H} = 10^{23.46}~cm^{-2}$,][]{2019Natur.573...83Z} was observed along the sightline with $\beta=45^\circ$, we can constrain that our infalling \MgII\ NALs ($\rm N_{\rm H} = 10^{20}~cm^{-2}$) were observed along the directions with $\beta\approx75^\circ$, which are quite close to the axes of accretion disks. Adopting $\Omega_{\rm inflow}=0.6$ \citep[][]{2019Natur.573...83Z}, we find that $\rm \dot{M}_{\rm inflow}$=(0.0110 --- 0.1487) $M_{\bigodot}/yr$ if the infalling \MgII\ NALs are formed within dusty tori, which are about one order of magnitude lower than the accretion rates of black holes (see the red stars in Figure \ref{fig:Mar}).

Alternatively, the infalling \MgII\ NALs may result from the interaction of AGN jets with the interstellar medium (ISM), which leads to cold gas precipitation and chaotic cold accretion onto central SMBHs \citep[e.g.,][]{2013MNRAS.432.3401G,2014ApJ...789..153L}. The jet-ISM interaction leads to nonlinear development of local thermal instability, and cold gas precipitates and falls back to the central regions preferentially along the directions close to the jet axis. Simulations indicate that with this mechanism, a majority of the cold gas forms at kiloparsec and larger distances away from the central SMBH. The detected infalling NALs are located on the parsec scale along the sightlines quite close to the axes of accretion disks, and therefore they may be originally formed nearly along the jet axes on the kiloparsec scale. A natural place to form these cold gaseous blobs is metal-rich trailing outflows uplifted by the jet ejecta through the Darwin drift mechanism \citep[][]{2018MNRAS.473.1332G,2018ApJ...861..106D}. The initially hot or warm gas converges and cools rapidly in trailing outflows, and the resulted cold NALs experience a transition from an initial outflowing stage to a later inflowing stage. The whole process is expected to last for about 100 Myr, long after the jet ejecta becomes undetectable at 20 cm. This picture is consistent with the rather high fraction (5/9) of the FIRST radio detection of our infalling \MgII\ NALs. The NAL absorbers may thus provide direct evidence for cold gas precipitation and accretion in AGN feedback processes. In this picture, we expect that $\Omega_{\rm inflow}$ is significantly lower than 1. Adopting $\Omega_{\rm inflow}=0.03$ that corresponds to a half-opening angle of $15^{\circ}$, we find that $\rm \dot{M}_{\rm inflow}$=(0.0005 --- 0.0074) $M_{\bigodot}/yr$, which is about 2 --- 3 orders of magnitude lower than the accretion rates of black holes (see the blue squares in Figure \ref{fig:Mar}). These low mass inflow rates are consistent with the suggestions that the chaotic cold accretion cannot provide sufficient gas to trigger luminous quasars \citep[e.g.,][]{2017ApJ...837..149G,2019NatAs...3...48S}.

In term of above discussions, it does not matter whether the infalling \MgII\ NALs are from the dusty tori or the chaotic cold accretion, the infalling NALs cannot provide sufficient fuels to power the quasars. Even if we adopt the upper limits of $C_{\rm f}=1$ and $\Omega_{\rm inflow}=1$, the mass inflow rate is still obviously less than the mass accretion rate of black hole (see the green triangles in Figure \ref{fig:Mar}).

The inflowing gas of BALs, or most of them, moves mainly along the directions close to the equatorial plane, where the radiation flux from the central engine is relatively small. The BAL absorbers will thus experience weak radiation pressure and would be pulled onto accretion disks by the gravity of central SMBHs and provide materials to feed the growth of central SMBHs. However, the inflowing gas of NALs move mainly along the directions close to the rotational axis of accretion disks. Strong radiation flux exists along these directions. Therefore, the NAL absorbers are expected to be decelerated by the radiation pressure. They may even become outflows by the radiation line force although the quasar radiation is sub-Eddington \citep[][]{2007ApJ...661..693P,2017ApJ...837..149G}. Therefore, the ultimate destiny of inflowing NALs is possibly different from that of inflowing BALs, where the inflowing BALs are transported to the accretion disks, and then to feed the SMBHs.

\section{Summary}
Using the large quasar catalog from the SDSS, this work aims to search for redshifted/infalling \MgII\ NALs, which is benefit to comprehend the feed and feedback mechanisms and processes of quasars. We obtain 9 robust \MgII\ NALs with $\upsilon_r<-1000$ \kms, which are identified by \MgIIab\ doublet, and at least two absorption lines at other rest-frame wavelengths, such as the series of absorption lines of \FeII, and the \CIVab. Every system is identified by more than four narrow absorption lines, which guarantees that our 9 \MgII\ NALs are reliable. Both the redshifts of the quasar ($z_{\rm em}$) and absorption ($z_{\rm abs}$) systems are determined by narrow lines, which guarantees the high accuracy of both the $z_{\rm em}$ and $z_{\rm abs}$. In addition, the large inflowing speed ($\upsilon_r<-1000$ \kms) ensures that our 9 \MgII\ NALs are much likely infalling towards quasar central regions. They are very unlikely to be originated in the external galaxies that are randomly moving in the cluster of quasar host galaxy, and much unlikely to be formed within the gas clouds located within the quasar host galaxy and CGM.

The 9 redshifted \MgII\ NALs are infalling towards quasar central regions in speeds of 1071 --- 1979 \kms, and located around the dusty tori (parsec scale from central SMBHs). Both the narrow line characteristic of absorption systems and strong radio emissions of quasars suggest that these redshifted \MgII\ NALs are likely along the directions close to the axes of accretion disks.

The quasars showing inflows might have lower radiation pressure when compared to the sources with outflows, which will be further studied when a large sample of quasars with confirmed outflow absorptions are available in the future.

Although we cannot confirm where the inflowing \MgII\ NALs come from with present data, there are two possible origins. One possible origin is that the inflowing \MgII\ NALs are formed within the dusty tori. The other one possibility is that the inflowing \MgII\ NALs are from the chaotic cold accretion resulted from the nonlinear interaction of AGN jets with the interstellar medium, where the cold gaseous blobs may originally precipitate in metal-rich trailing outflows uplifted by AGN jet ejecta. Therefore, our inflowing \MgII\ NALs might be the direct evidence of cold gas precipitation and accretion in AGN feedback processes, and the direct evidence of inflowing gas along the directions close to quasar jets and at parsec scale. Of course, we note that the mass inflow rates of the infalling \MgII\ NALs are much less than the mass accretion rates of black holes. Thus the infalling NALs cannot provide sufficient fuels to power the quasars.

\begin{acknowledgements} Zhi-Fu Chen is supported by the Guangxi Natural Science Foundation (2019GXNSFFA245008), National Natural Science Foundation of China (12073007), the Guangxi Natural Science Foundation(GKAD19245136; 2018GXNSFAA050001), and Scientific research project of Guangxi University for Nationalities (2018KJQD01), National Natural Science Foundation of China (11763001). MFG is supported by the National Science Foundation of China (grant 11873073). Zhi-Cheng He is supported by NSFC-11903031, NSFC 12192221, and USTC Research Funds of the Double First-Class Initiative YD 3440002001. Fu-Lai Guo is supported by National Natural Science Foundation of China (11873072).
\end{acknowledgements}


\bibliographystyle{aa}
\bibliography{ms}

\end{document}